\documentclass[10pt]{iopart}

\usepackage{graphicx}
\usepackage{morefloats}
\usepackage{setstack}
\usepackage{color}
\usepackage{epsf,epsfig,graphics}

\begin{document}

\title[DR rate coefficients for W$^{55+}$ to W$^{38+}$]{Partial and Total Dielectronic Recombination Rate Coefficients 
for W$^{55+}$ to W$^{38+}$}

\author{S. P. Preval}\author{N. R. Badnell}\author{M. G. O'Mullane}

\address{Department of Physics, University of Strathclyde, Glasgow G4 0NG, United Kingdom}
\ead{simon.preval@strath.ac.uk}
\vspace{10pt}
\begin{indented}
\item[]November 2016
\end{indented}

\begin{abstract}
Dielectronic recombination (DR) is the dominant mode of recombination in magnetically confined fusion plasmas for
intermediate to low-charged ions of W. Complete, final-state resolved partial isonuclear W DR rate coefficient data is required for
detailed collisional-radiative modelling for such plasmas in preparation for the upcoming fusion experiment ITER. To
realize this requirement, we continue {\it The Tungsten Project} by presenting our calculations for tungsten
ions W$^{55+}$ to W$^{38+}$. As per our prior calculations for W$^{73+}$ to W$^{56+}$, we use the collision package {\sc
autostructure} to calculate partial and total DR rate coefficients for all relevant core-excitations in intermediate
coupling (IC) and configuration average (CA) using $\kappa$-averaged relativistic wavefunctions. Radiative recombination
(RR) rate coefficients are also calculated for the purpose of evaluating ionization fractions. Comparison of our
DR rate coefficients for W$^{46+}$ with other authors yields agreement to within 7-19\% at peak abundance verifying the
reliability of our method. Comparison of partial DR rate coefficients calculated in IC and CA yield differences of
a factor $\sim{2}$ at peak abundance temperature, highlighting the importance of relativistic configuration mixing. Large
differences are observed between ionization fractions calculated using our recombination rate coefficient data and
that of P\"{u}tterich~\etal [Plasma Phys. and Control. Fusion 50 085016, (2008)]. These differences are attributed to 
deficiencies in the average-atom method used by the former to calculate their data.
\end{abstract}

%
%
\submitto{\JPB}
%
%
\ioptwocol

\section{Introduction}
The upcoming experimental fusion reactor ITER \footnote{{h}ttp://www.iter.org} is currently being constructed in
Cadarache, France. Scheduled for a first plasma in December 2025\footnote{{h}ttps://www.iter.org/newsline/-/2482}, ITER
has been designed with the aim of producing ten times as much energy as it consumes for operations. The reactor wall of
ITER will be constructed with beryllium tiles, and the divertor will be composed of tungsten \cite{raffray2014a}. In
preparation for the experiments that will take place at ITER, the Joint European Torus (JET) based in Culham, Oxford,
has been fitted with an ITER-like wall \cite{matthews2011a}. In addition, the ASDEX upgrade \cite{neu2007a} now uses
nearly 100\% tungsten in its set-up. Tungsten has been chosen for its ability to withstand large power loads even in the
presence of micro-fractures, resistance to tritium absorption \cite{kharbachi2014a}, and high melting point. However,
being a plasma facing component, tungsten will sputter from the divertor into the main body plasma, potentially cooling
and quenching the plasma. The power loss \cite{summers1979a} arising from the presence of tungsten impurities can be
calculated using sophisticated collisional-radiative (CR) models \cite{summers2006a}. Partial final-state resolved DR rate
coefficients are required not just for high-$n$ but also for population modelling of low-lying energy levels. As input,
CR models require accurate isonuclear partial final-state resolved dielectronic recombination (DR) rate coefficient
data for the elements being included.

\begin{figure}
\begin{centering}
\includegraphics[width=85mm]{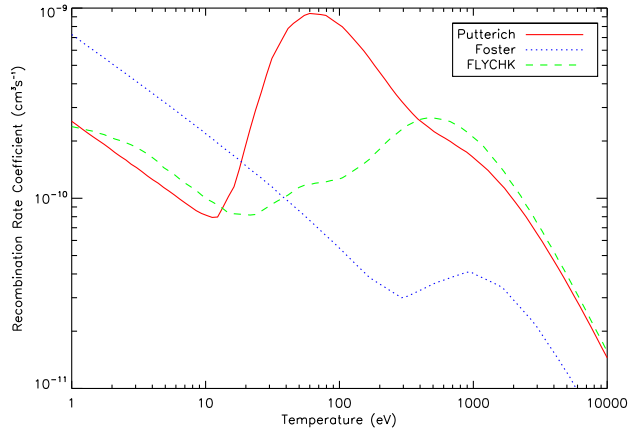}
\caption{Total recombination rate coefficients for W$^{44+}$ as calculated by P\"{u}tterich~\etal (solid red) \cite{putterich2008a}, Foster (dotted 
blue) \cite{foster2008a} and FLYCHK (dashed green). The FLYCHK data location is given in text.}
\label{fig:intropl}
\end{centering}
\end{figure}

Several calculations have been performed in response to the demand for tungsten DR rate coefficient data. The first
attempt at a baseline set of data was done by \cite{post1977a,post1995a} using an average-atom method, ADPAK. 27 years
later, tungsten DR rate coefficient data\footnote{{h}ttps://www-amdis.iaea.org/FLYCHK/ZBAR/csd074.php} was calculated
using the relativistic configuration average (CA) code FLYCHK \cite{chung2005a}. Next, Foster \cite{foster2008a}
used the Burgess General Formula \cite{burgess1965a} to calculate data for tungsten. Finally, P\"{u}tterich~\etal
\cite{putterich2008a} used the data of ADPAK, but multiplied the DR rate coefficients of several ionization stages
(W$^{22+}$ - W$^{55+}$) by empirical scaling factors to alter the theoretical ionization balance and improve agreement
with observed spectral emission. Both Foster and P\"{u}tterich~\etal used scaled hydrogenic radiative recombination
(RR) rate coefficients. While these approaches cover the isonuclear sequence, agreement between the three methods
is lacking. For example, in Figure \ref{fig:intropl} we show the recombination rate coefficients for W$^{44+}$ as
calculated by P\"{u}tterich~\etal, Foster, and Chung~\etal. It can be seen there is little-to-no similarity between
the results of the three calculations. At the very least, the variation between the three curves give an indication
on the uncertainty on the data, however, this is not very useful for calculating detailed CR models.

Large uncertainties in calculated recombination rate coefficients translate to uncertainties in peak abundance
temperatures. In figure 1 of \cite{preval2016a} we plotted the tungsten ionization balance calculated using the
ionization rate coefficients of Loch~\etal \cite{loch2005a}, and the recombination rate coefficients of Foster and
P\"{u}tterich~\etal. It was shown that while there is moderate agreement for the ten highest ionization stages where
RR is dominant, the position of and size of the fractions for each ionization stage differs markedly.

Calculating final-state resolved partial DR rate coefficients is a more involved process, and has only been done for
a select few ionization stages of tungsten. Typically, the ionization stages considered have been for simpler cases
where the electron shell is near- or completely filled, however, there are exceptions. To date, the most complex
tungsten ions to have been studied are the open-f shell ions W$^{18+}$, W$^{19+}$, and W$^{20+}$ ($4d^{10}4f^{q}$,
$q=8,9,10$) by Spruck~\etal and Badnell~\etal \cite{spruck2014a,badnell2016a,badnell2012a} using an updated version
of {\sc autostructure} to cope with the large number of levels. A less computationally demanding set of calculations
were carried out by Preval~\etal \cite{preval2016a}, who also used {\sc autostructure} to calculate partial and total
DR rate coefficients for W$^{73+}$ to W$^{56+}$ as part of {\it The Tungsten Project}. The RR rate coefficients were
also calculated alongside these for use in ionization balance comparisons.

In addition to our calculations using {\sc autostructure}, other codes have been used to calculate
detailed DR rate coefficients for Tungsten. Behar~\etal used the {\sc hullac} code \cite{barshalom2001a}
and the Cowan code \cite{cowanbook1981} to calculate DR rate coefficients for W$^{46+}$, W$^{45+}$, W$^{56+}$
\cite{behar1997a,behar1999b}, and W$^{64+}$ \cite{behar1999a}. Peleg~\etal \cite{peleg1998a} also used {\sc hullac}
and the Cowan Code to calculate DR rate coefficients for W$^{56+}$. In addition, Safronova~\etal has used {\sc hullac}
to calculate DR rate coefficients for W$^{4+}$, W$^{6+}$, W$^{28+}$, W$^{45+}$, W$^{46+}$, W$^{63+}$, and W$^{64+}$
\cite{usafronova2012a,usafronova2012b,usafronova2011a,usafronova2015a,usafronova2012c,usafronova2009a,usafronova2009b}. The
Flexible Atomic Code ({\sc fac}) \cite{gu2003a} was used by Li~\etal to calculate DR rate coefficients for W$^{29+}$,
W$^{39+}$, W$^{27+}$, W$^{28+}$, and W$^{64+}$ \cite{li2012a,li2014a,li2016a}. Meng et al. and Wu et al. also used {\sc fac}
to calculated data for W$^{47+}$, and W$^{37+}$ - W$^{46+}$ respectively \cite{meng2009a} \cite{wu2015a,wu2015b}. More
recently, Kwon et al. calculated DR rate coefficients for W$^{45+}$ \cite{kwon2016a}. However, these authors noted that
important channels were left out of their calculation. Kwon et al later published an erratum including the updated
calculation, as well as the rate coefficients for two additional ions, W$^{46+}$ and W$^{44+}$ \cite{kwon2016b}. All
of these codes are distorted wave. Ballance~\etal used the Dirac Atomic R-Matrix ({\sc darc}) to calculate DR rate
coefficients for W$^{35+}$ \cite{ballance2010a} to provide a comparison with CA and distorted wave methods.

{\it The Tungsten Project} was described by Preval~\etal, which is a programme of work in which we aim to calculate
partial final-state resolved and total zero-density DR rate coefficients for the entire isonuclear sequence of
tungsten. The calculated data from the project is hosted on the Open Atomic Data and Analysis Structure (OPEN-ADAS)
website \footnote{{h}ttp://open.adas.ac.uk} in the standard {\it adf09} (DR) and {\it adf48} (RR) formats, the
definitions of which can be found on \footnote{{h}ttp://www.adas.ac.uk}.

As introduced by Preval~\etal, we use some technical notation when referring to the various ionization stages of
tungsten. As it is not particularly helpful to refer to elements with atomic number $Z>30$ by their isoelectronic
symbols, we refer to different ionization states by the number of their valence electrons. In this scheme, H-like
($Z=1$) becomes 01-like, Zn-like ($Z=30$) becomes 30-like, and Pa-like ($Z=46$) becomes 46-like, and so on.

In this paper we consider the ionization stages 19- to 36-like, which covers ions with ground configurations $3d^{q}$
($q=1,10$), and $4s^{r}$ $4p^{t}$ ($r=1,2$ and $t=0-6$). The paper is structured as follows. Section \ref{theorysec}
gives a brief overview of the theory used in calculating the DR/RR rate coefficients. In Section \ref{calcsec},
we discuss the calculations performed for DR and RR, including the configurations used, and the core-excitations
considered. In Section \ref{resultsec} we discuss the results obtained, and consider the effects of relativistic
configuration mixing on the partial and total DR rate coefficients. We make comparisons between our calculated data,
and those mentioned in the previous paragraphs. We also quantify changes to the ionization balance when using the
newly calculated data. We then summarise our results and make some concluding remarks.
 
\section{Theory}\label{theorysec}
The underlying theory of our calculations was discussed at length in Preval~\etal \cite{preval2016a}; however,
we provide a brief recap here. The DR rate coefficients were calculated using the distorted-wave collision package
{\sc autostructure} \cite{badnell1986a,badnell1997a,badnell2011a}, which uses the independent processes and isolated
resonances approximation \cite{pindzola1992a}. {\sc autostructure} has been tested at length by comparing with theory
(see Savin~\etal \cite{savin2005a}) and experiment over the past 30 years. The partial DR recombination rate coefficient
$^{DR}\alpha_{f\nu}^{z+1}$, from initial state $\nu$ of ion $X^{+z+1}$ to a final state $f$ of ion $X^{+z}$, can be
written as
\begin{eqnarray}
^{DR}\alpha_{f\nu}^{z+1}(T_e)&=&\left(\frac{4{\pi}a_{0}^{2}I_{H}}{k_{B}T_{e}}\right)^{\frac{3}{2}}
                                \sum_{j}\frac{\omega_{j}}{2\omega_{\nu}}\exp{\left[-\frac{E}{k_{B}T_{e}}\right]} \nonumber \\
                                &\times&\frac{\sum_{l}{A_{j\rightarrow{\nu},E\,l}^a}{A_{j\rightarrow{f}}^{r}}}
                                {\sum_{h}{A_{j\rightarrow{h}}^{r}} + \sum_{m,l}{A_{j\rightarrow{m},E\,l}^{a}}},
\label{DReq}
\end{eqnarray}
where the $A^{a}$ are the autoionization rates, $A^{r}$ are the radiative rates, $\omega_{\nu}$ 
is the statistical weight of the $N$-electron target ion, $\omega_{\j}$ is the statistical weight of the
$(N+1)$-electron state, $E$ is the total energy of the 
continuum electron, minus its rest energy, and with corresponding orbital angular momentum 
quantum number $l$ labelling said channels. The sum over $j$ represents the sum over all
autoionizing states. The sum over $h$ and $m$ represent the total radiative and autoionization widths ($\hbar=1$) respectively.
$I_{H}$ is the ionization energy of the 
hydrogen atom, $k_{B}$ is the Boltzmann constant, $T_{e}$ is the electron temperature, and 
$(4{\pi}a_{0}^{2})^{3/2}=6.6011\times{10}^{-24}$cm$^{3}$.

While no longer dominant, the RR rate coefficients are still significant, contributing ${14}$\% to the total recombination rate coefficient 
for 19-like. The total recombination rate is necessary to calculate the ionization balance. Using detailed balance,
the partial RR rate coefficient $^{RR}\alpha_{f\nu}^{z+1}(T_e)$ can be written in terms of its inverse process, photoionization $^{PI}\sigma_{\nu f}^{z}(E)$:
\begin{eqnarray}
^{RR}\alpha_{f\nu}^{z+1}(T_e) &=& \frac{c\,\alpha^3}{\sqrt{\pi}}
                                  \frac{\omega_{f}}{2\omega_{\nu}}\left(I_H k_B T_e\right)^{-3/2} \nonumber \\
                                  &\times&\int^\infty_0 E^2_{\nu f}\, {^{PI}\sigma_{\nu f}^{z}(E)}
                                  \exp{\left[-\frac{E}{k_{B}T_{e}}\right]}dE\,,
\label{RReq}
\end{eqnarray}
where $E_{\nu f}$ is the corresponding photon energy and $c\,\alpha^3/\sqrt{\pi}=6572.67$cm\,s$^{-1}$.

Relativistic corrections to the Maxwell-Boltzmann distribution become important at high temperatures. This correction manifests as a multiplicative factor to give the
Maxwell-J\"{u}ttner distribution \cite{synge57a} as 
\begin{equation}
F_{\mathrm{r}}(\theta) = \sqrt{\frac{\pi\theta}{2}} 
\frac{1}{K_{2}(1/\theta){\rm e}^{1/\theta}},
\end{equation}
where $\theta=\alpha^2 k_{B}T/2I_H$, $\alpha$ is the fine-structure constant and $K_{2}$ is the modified Bessel function of 
the second kind. 

\section{Calculations}\label{calcsec}

\subsection{DR}

We split our calculations into core-excitations. These core-excitations are labelled by the initial and final principal quantum
numbers $n$ and $n'$ of the promoted target electron. Excitations into the various $\ell$-values associated with $n$ and $n'$ is implied with this notation.
The total contribution from inner-shell DR ($n=n'-1$, $n=2$) for 19- to 27-like is relatively small, as the full $3s$ and $3p$ shells restricts
excitations only to $3d$. Comparatively, the contribution from inner-shell DR ($n=n'-1$, $n=3$) for 29-like onwards is much
larger, as there are up to two 3s and six 3p electrons available to promote to $4\ell$. Core-excitations 
where $\Delta{n}>1$ tend to give smaller contributions to the total rate coefficients ($<10$\%), as DR tends to be suppressed by 
autoionization into excited states. It is for this reason that we opt to calculate the inner-shell DR ($n=n'-1$, $n=2$) and 
$\Delta{n}>1$ core-excitations in CA. 

A full list of the core-excitations we calculated DR rate coefficients for is given in Table \ref{table:drcorex}. For 
outer-shell DR, the $N$-electron configurations included consist of all possible single-electron promotions from $nl$ to $n'\ell'$, where $n$ and $n'$ are given
by the core-excitation being considered. In addition, configurations accessible by mixing are included.
Mixing effects are strongest for terms/levels that are close in energy. The strongest mixing 
configurations can be determined using a ``one-up, one-down" rule. For example, in the $n=4$ complex, if we have a 
configuration $4s^{\mu} 4p^{\nu} 4d^{\rho} 4f^{\sigma}$, where $\mu$, $\nu$, $\rho$,and $\sigma$ are occupation numbers,
then the strongest mixing configurations expected would be:
\begin{itemize}
\item $4s^{\mu-1} 4p^{\nu+1} 4d^{\rho+1} 4f^{\sigma-1}$
\item $4s^{\mu+1} 4p^{\nu-1} 4d^{\rho-1} 4f^{\sigma+1}$
\item $4s^{\mu-1} 4p^{\nu+2} 4d^{\rho-1} 4f^{\sigma}$
\item $4s^{\mu+1} 4p^{\nu-2} 4d^{\rho+1} 4f^{\sigma}$
\item $4s^{\mu} 4p^{\nu-1} 4d^{\rho+2} 4f^{\sigma-1}$
\item $4s^{\mu} 4p^{\nu+1} 4d^{\rho-2} 4f^{\sigma+1}$.
\end{itemize}
The $(N+1)$-electron configurations included are simply the $N$-electron configurations with an additional target electron added to the 
final recombined $n'$ shell. As an example, we give a list of configurations included for the 30-like 4--5 core-excitation calculation
in Table \ref{table:drexam1}. For inner-shell DR, the configurations included are similar to that of outer-shell DR, except additional
core-rearrangement configurations are included. We include an example of the configuration set for 30-like 3-4 in Table
\ref{table:drexam2}.

For the Rydberg $n\ell$ electron calculation, we calculate radiative/autoionization rates for all individual $n$ up to $n=25$, 
after which they were calculated for logarithmically spaced $n$ up to $n=999$. This was done up to $\ell$-values such that 
the total DR rate coefficients for a particular core-excitation are converged numerically to $<1$\% over the entire ADAS 
temperature range, given by $(10 - 10^{6})z^2$K, where $z$ is the residual charge of the target ion.

As one moves deeper into the $3d$ shell the complexity of the calculation increases dramatically. It quickly becomes necessary to 
neglect certain subshells and restrict electron promotions when calculating DR rate coefficients in IC. Fortunately, only the 3--4 
core-excitation needs to be restricted in this way. For ions 20-like to 33-like promotions from $3s\rightarrow{4\ell}$ can be 
neglected. For ions 34- to 36-like promotions from $3s,3p\rightarrow{4\ell}$ can be neglected. 

We performed calculations in CA to estimate the contributions of promotions from these neglected subshells in IC to the total DR 
rate coefficient for the 3--4 core-excitation. In the case of 20-like, the $3s\rightarrow{4\ell}$ promotions contribute 
9\% to the total 3--4 core-excitation at peak abundance ($1.08\times{10}^{8}$K) for 20-like. Overall, the 3--4 core-excitation contributes 
52\% to the total recombination rate coefficient, resulting in a 5\% difference to the totals upon neglecting the $3s$ contribution. 
The $3s\rightarrow{4\ell}$ contribution decreases rapidly as the $3d$, $4s$, and $4p$ shells are filled. By 28-like, 
$3s\rightarrow{4\ell}$ promotions contribute 1\% to the total at peak abundance. In the case of 33- to 36-like, the 
$3p$ contribution to the total 3--4 core-excitation for 33-like is just 3\% at peak abundance, again decreasing rapidly to $<1$\% by 36-like.
For 34-like onwards, we also close the $3p$ shell, considering only $3d-4\ell$ contributions.

\subsection{RR}
RR is simpler to calculate in comparison to DR. Preval~\etal \cite{preval2016a} showed that RR was very important at the highest ionization
stages. The set of configurations included mirrored the $\Delta{n}=0$ DR configurations for simplicity. In reality, the set of configurations that 
need to be included in an RR calculation is smaller than this. Including additional configurations increases the computational time,
and contributes little to the final result. Therefore, the $N$-electron configurations retained consist of the 
ground plus single excitations of the outermost electron within the complex. We also include any mixing configurations 
as given in the previous subsection. The $(N+1)$-electron configurations are just the $N$-electron configurations with an extra target electron.
As an example, we have listed the $N$- and $(N+1)$-electron configurations included for 30-like in Table \ref{table:rrexam}.

RR rate coefficients were calculated for all individual $n$ up to $n=25$, after which they were calculated for logarithmically spaced 
$n$ up to $n=999$. We included $\ell$-values up to $\ell=10$ relativistically, after which a non-relativistic, hydrogenic ''top-up" 
was included to numerically converge the total to $<1$\% over the entire ADAS temperature range. The RR rate coefficients were 
calculated in both IC and CA. In the IC case, contributions from multipolar radiation up to E40 and M39 were included, whereas 
for CA, multipolar radiation contributions up to E40 alone were included.

\section{DR - Results}\label{resultsec}
In this section we describe the DR rate coefficients calculated. We opt to split this section up by the complex being filled, 
and the core-excitation being considered.

\subsection{DR of $3d^q$, $q=1,10$}
\begin{figure}
\begin{centering}
\includegraphics[width=85mm]{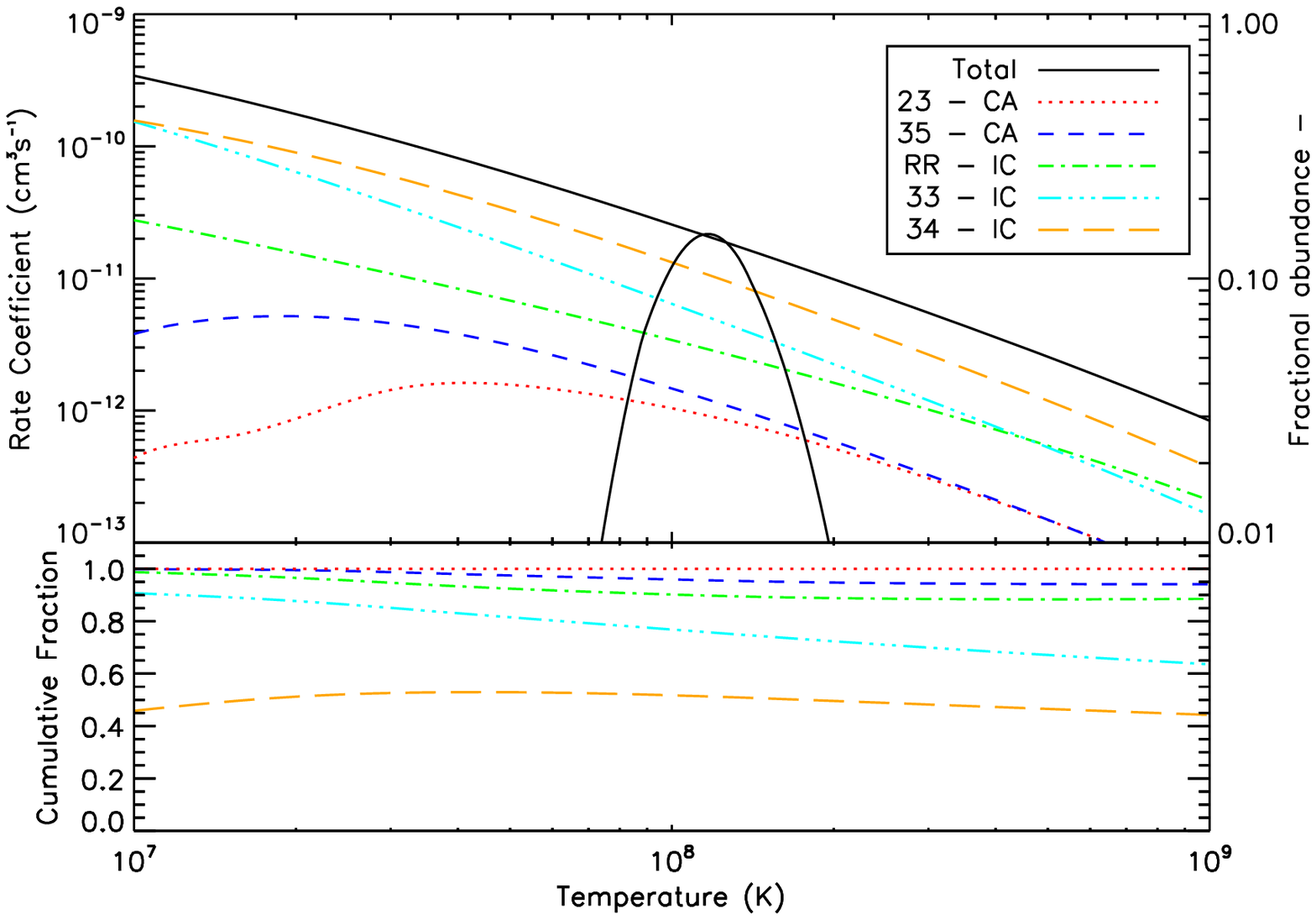}
\caption{DR rate coefficients for each core-excitation of 19-like tungsten, the RR rate coefficient, and the total of
these. The parabola is the ionization fraction for 19-like, calculated using P\"{u}tterich~\etal \protect\cite{putterich2008a}'s 
recombination rate coefficients, and Loch~\etal \protect\cite{loch2005a}'s ionization rate coefficients. The bottom plot gives the 
cumulative contribution of each recombination rate coeffiicent, and is described in text.}
\label{fig:klike}
\end{centering}
\end{figure}

\begin{figure}
\begin{centering}
\includegraphics[width=85mm]{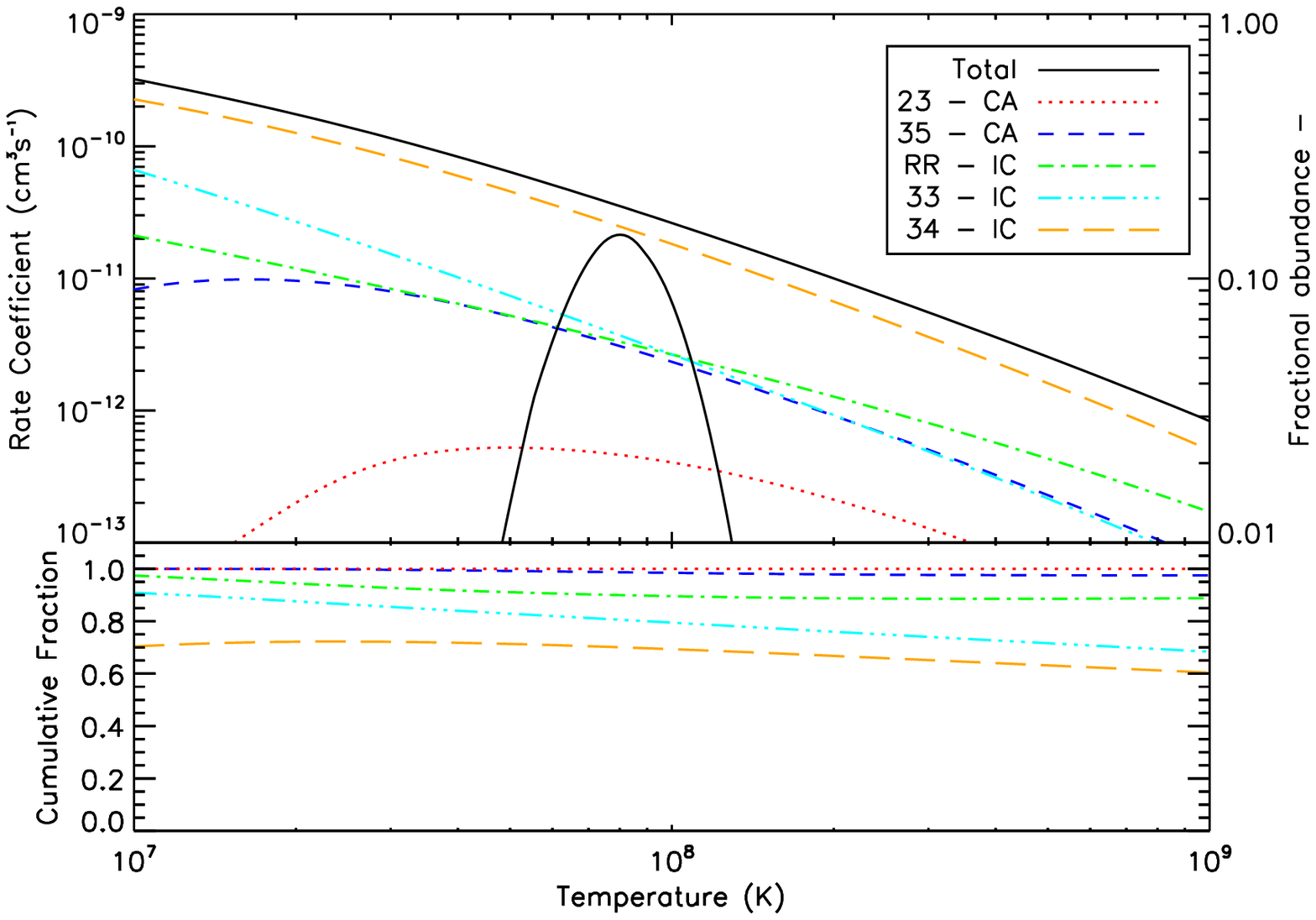}
\caption{The same as Figure \ref{fig:klike}, but for 24-like.}
\label{fig:mnlike}
\end{centering}
\end{figure}

The $3d^q$ ions cover 19-like to 28-like. 28-like, or Ni-like, with a $3d^{10}$ ground state. It is an important ionization stage where the relatively few strong
excitation lines, because of its simplicity, are used in plasma diagnostics. In
Figures \ref{fig:klike} and \ref{fig:mnlike} we have plotted the total DR rate coefficients for 19- and 24-like tungsten along with 
the cumulative fraction of the individual contributions to the total. The cumulative fraction is calculated by adding successive 
core-excitations together, starting with the largest, and dividing the result by the total recombination rate coefficient. This allows 
a quantitative assessment of which core-excitations are important, and also shows the convergence of the total recombination rate
coefficient.

\subsubsection{2--3}
\begin{figure}
\begin{centering}
\includegraphics[width=85mm]{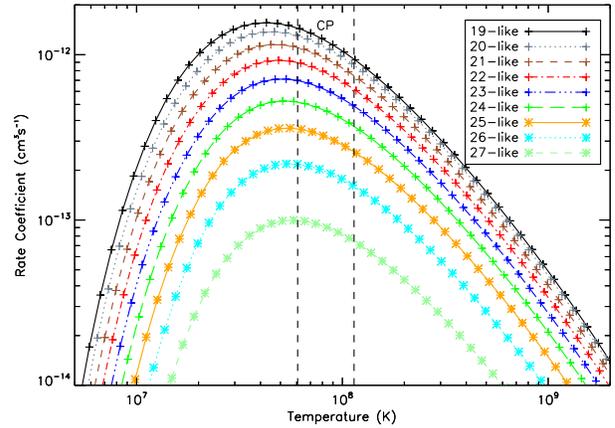}
\caption{Total DR rate coefficients for the 2--3 core-excitations for ionization stages 19- to 27-like tungsten, 
calculated in CA. The vertical dashed lines indicate the range of peak abundance temperatures from 19- to 27-like.}
\label{fig:23rates}
\end{centering}
\end{figure}

The 2--3 core-excitation provides the smallest contribution to the total DR rate coefficient. We include 2--3 for completeness,
and calculate this core-excitation in CA only due to its small contribution to the total DR rate coefficient. The 2--3 core-excitation contributes 
4\% to the total DR rate coefficient for 19-like at peak abundance temperature ($1.2\times{10}^{8}$K), dropping to 0.2\% at 27-like. In Figure \ref{fig:23rates} 
we plot the 2--3 DR rate coefficients for 19- to 27-like calculated in CA. In this plot, a region enclosed by two vertical dashed lines and 
labelled ``CP'' (collisionally ionized plasma) indicates the range of peak abundance temperatures from 19-like to 27-like. Note that the actual range of temperatures 
of interest will be wider than this.
It can be seen that the contribution from this core-excitation decreases rapidly as the residual charge decreases, due to the 3d shell being filled.

\subsubsection{3--3}
\begin{figure}
\begin{centering}
\includegraphics[width=85mm]{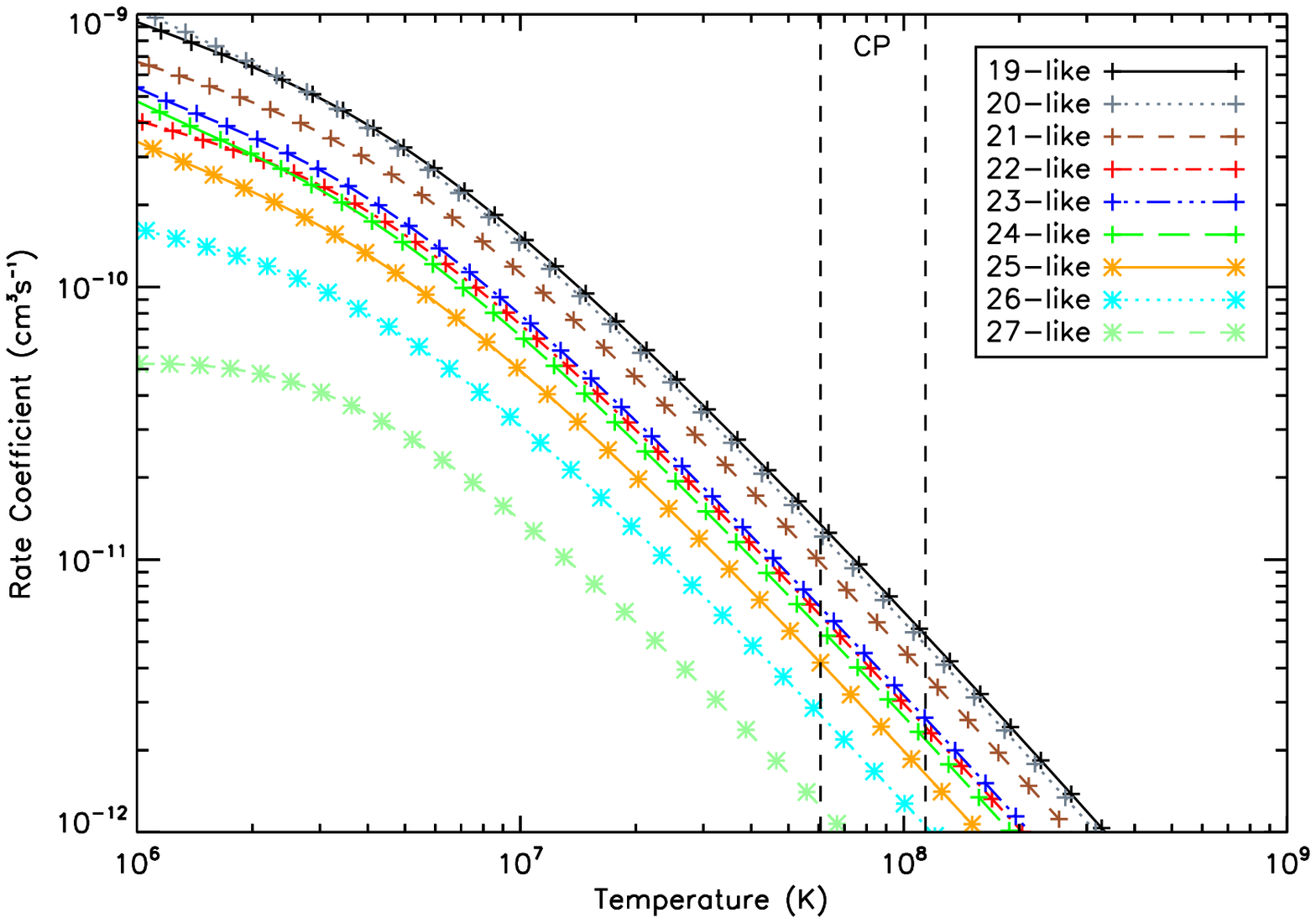}
\caption{Total DR rate coefficients for the 3--3 core-excitations for ionization stages 19- to 27-like tungsten, calculated in IC.}
\label{fig:33ratesic}
\end{centering}
\end{figure}

\begin{figure}
\begin{centering}
\includegraphics[width=85mm]{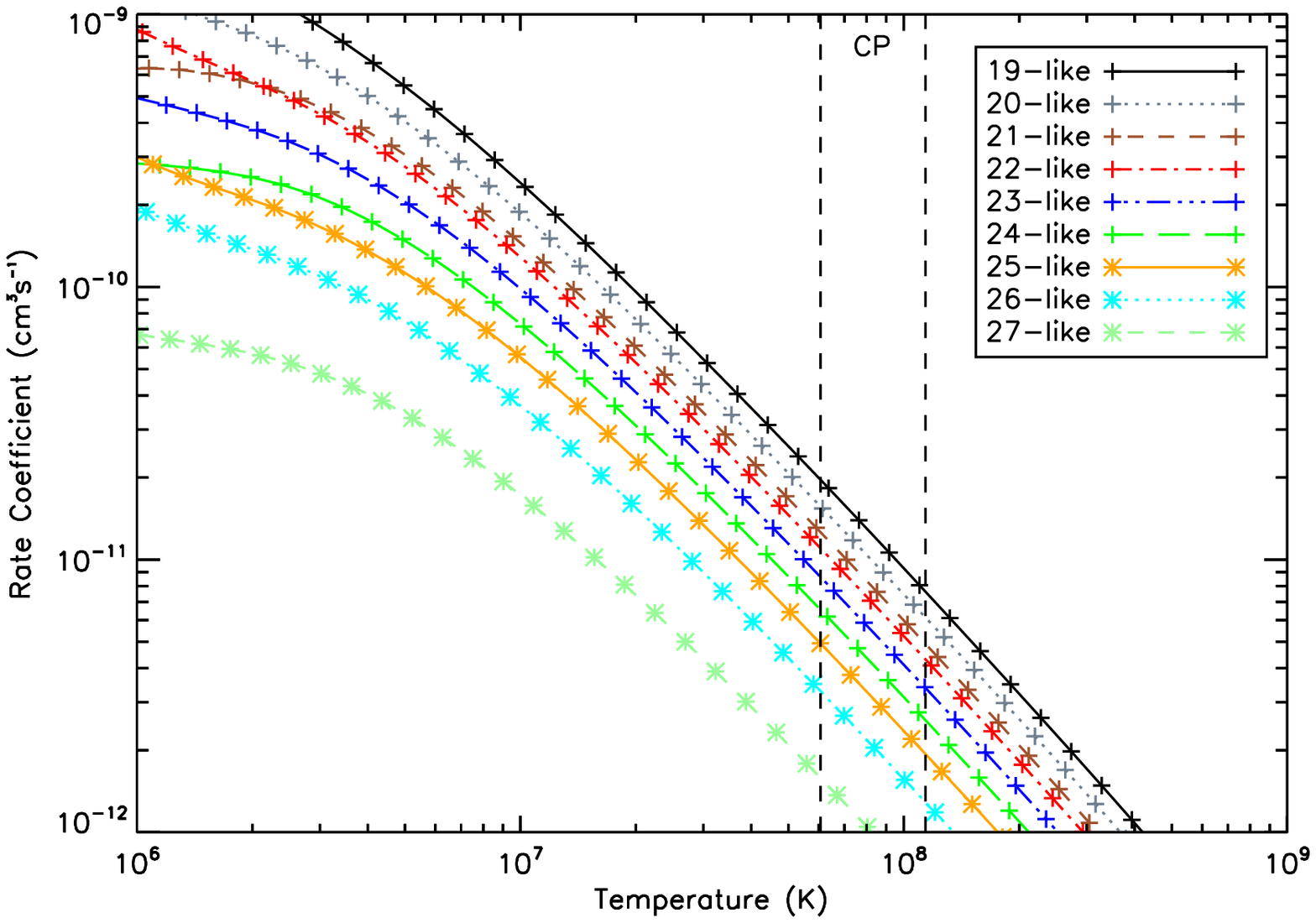}
\caption{The same as Figure \ref{fig:33ratesic}, but calculated in CA.}
\label{fig:33ratesca}
\end{centering}
\end{figure}

The 3--3 core-excitation provides a large contribution to the total DR rate coefficient at higher charges. Like 2--3,
the 3--3 contribution decreases steadily with decreasing residual charge. For 19-like, 3--3 contributes 24\% to the total DR rate coefficient 
at peak abundance, decreasing gradually to 3\% by 27-like. In Figures \ref{fig:33ratesic} and \ref{fig:33ratesca} we plot the 3--3 DR rate coefficients for 19- to 27-like in both IC
and CA, and indicate the CP range for these ions. There is not much difference between the total 3--3 DR rate coefficients when looking at the CA or IC results. In both 
cases, the rate coefficients decrease with decreasing residual charge. However, for the IC results, it can be
seen that 22- and 23-like swap order, with 23-like becoming larger than 22-like in the CP range. This suggests that 23-like is 
quite sensitive to relativistic configuration mixing, which is not present in the CA calculation.

\subsubsection{3--4}
\begin{figure}
\begin{centering}
\includegraphics[width=85mm]{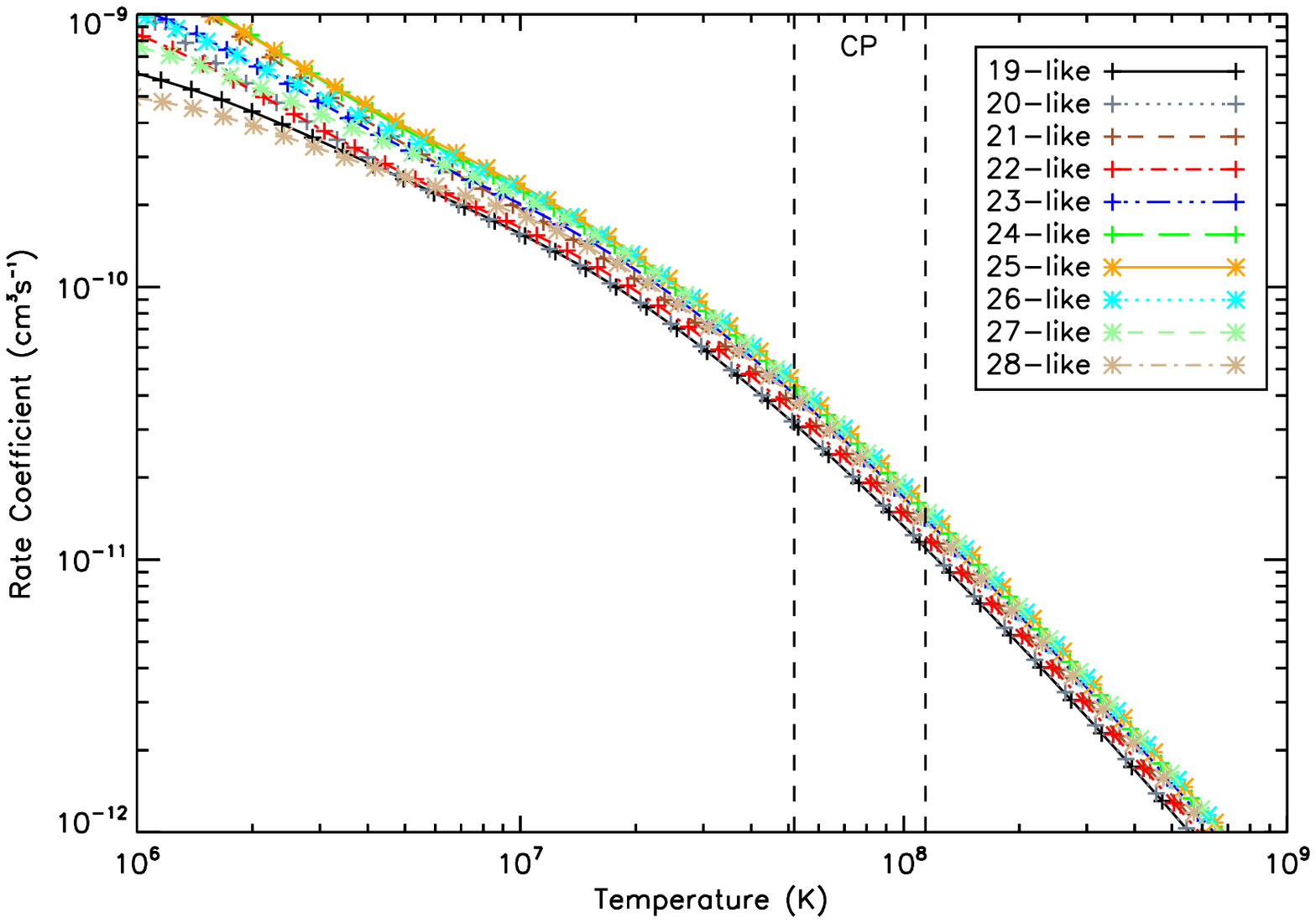}
\caption{Total DR rate coefficients for the 3--4 core-excitation for ionization stages 19- to 28-like tungsten, calculated in IC.}
\label{fig:34rates1ic}
\end{centering}
\end{figure}

\begin{figure}
\begin{centering}
\includegraphics[width=85mm]{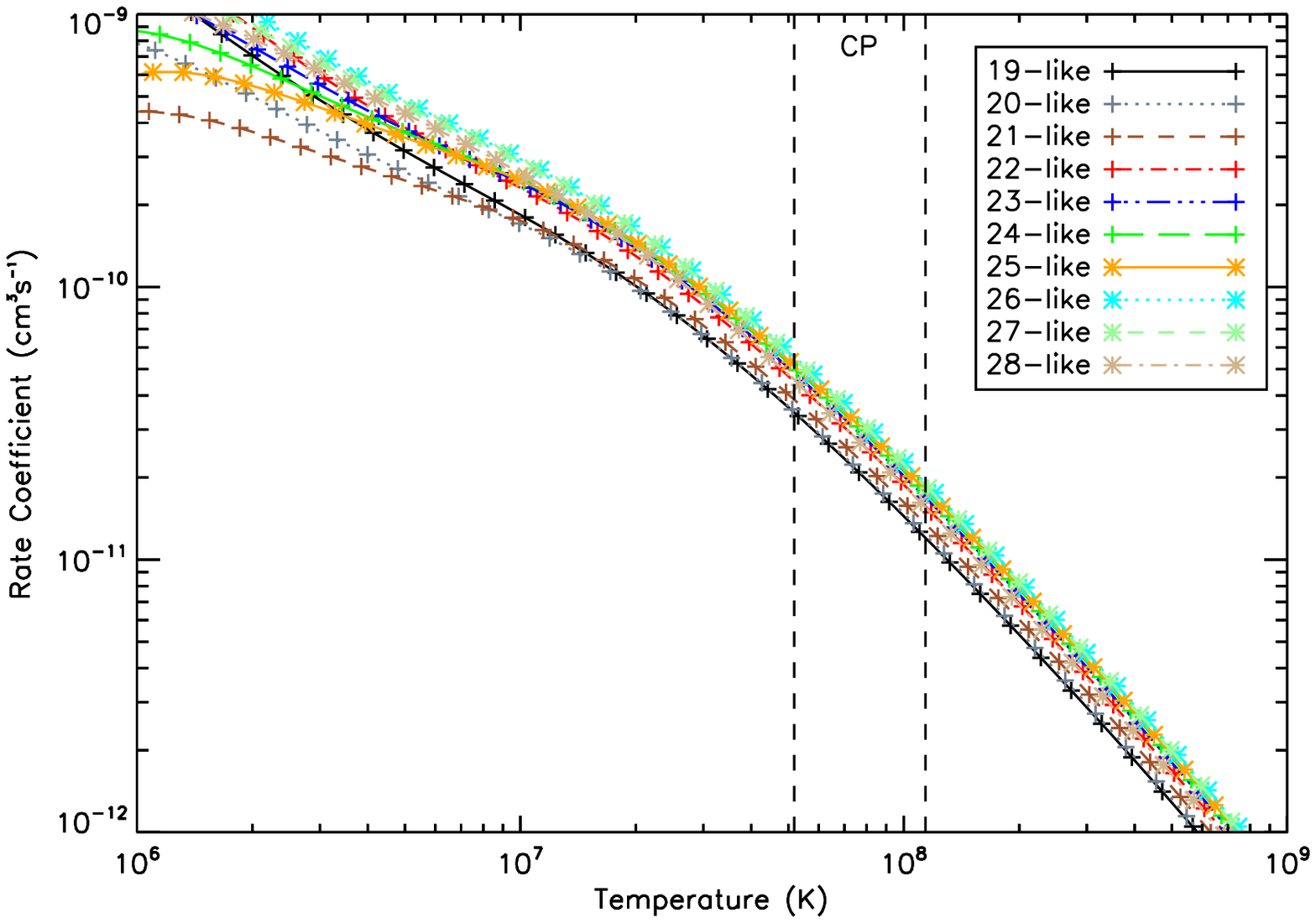}
\caption{The same as Figure \ref{fig:34rates1ic}, but calculated in CA.}
\label{fig:34rates1ca}
\end{centering}
\end{figure}

The 3--4 core-excitation provides by far the largest contribution to the total DR rate coefficient. This is due to the
large number of possible transitions that can take place by the six 3p electrons to 4$\ell$. The 3--4 core-excitation contributes 51\%
to the total DR rate coefficient for 19-like, increasing to 79\% by 28-like. In Figures \ref{fig:34rates1ic}
and \ref{fig:34rates1ca} we have plotted the 3--4 DR rate coefficients for 19-like to 28-like calculated in IC and
CA. We have also plotted the CP temperature region for these ions. As with 3--3, the difference between the IC and CA totals are
not very different. However, it can be seen that the 3--4 rates span a very narrow range in the CP region regardless
of the residual charge. This suggests that the DR rate coefficients for 3--4 are quite insensitive to changes in
the atomic structure. This insensitivity also gives an indication of the uncertainty of the calculated DR rate
coefficients. Furthermore, there does not appear to be any correlation between the DR rate coefficient and the residual charge. 
For example, instead of the DR rate coefficient increasing or decreasing in tandem with the residual charge,
some rate coefficients are higher than others. There are resonances close to threshold, and their positioning will cause any variation
seen at lower temperatures.

\subsubsection{3--5}
\begin{figure}
\begin{centering}
\includegraphics[width=85mm]{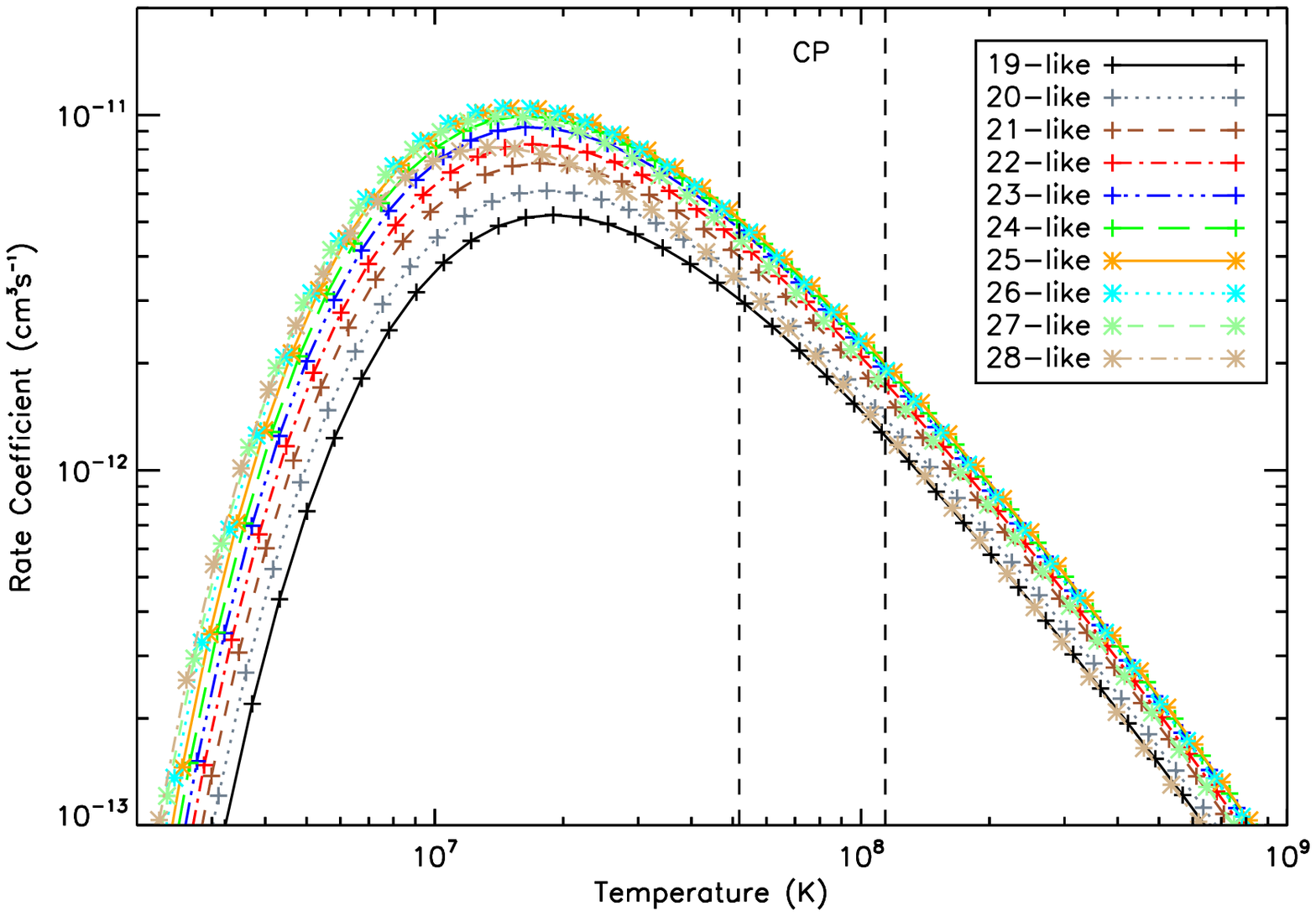}
\caption{Total DR rate coefficients for the 3--5 core-excitation for ionization stages 19- to 28-like tungsten, calculated in CA.}
\label{fig:35rates}
\end{centering}
\end{figure}

Like 2--3, the 3--5 core-excitation has been included for completeness, and has been calculated in CA only due to its small contribution
to the total DR rate coefficient. The 3--5 core-excitation contributes 6\% at peak abundance for 19-like. As we progress along 19- to 27-like, the 3--5 contribution
increases very gradually to 13\% by 28-like. In Figure \ref{fig:35rates} we have plotted the 3--5 DR rate coefficients for 19- to 28-like calculated in CA, and have indicated 
the CP temperature region for these ions. Like 3--4, it can be seen that the 3--5 DR rate coefficients are confined to a narrow band, and do not vary much with 
respect to ionization stage. This again indicates that the DR rate coefficients are relatively insensitive to the atomic structure. There are 
no resonances near threshold.

\subsection{DR of $4s^q$, $q=1,2$, and $4p^r$, $r=1,6$}
\begin{figure}
\begin{centering}
\includegraphics[width=85mm]{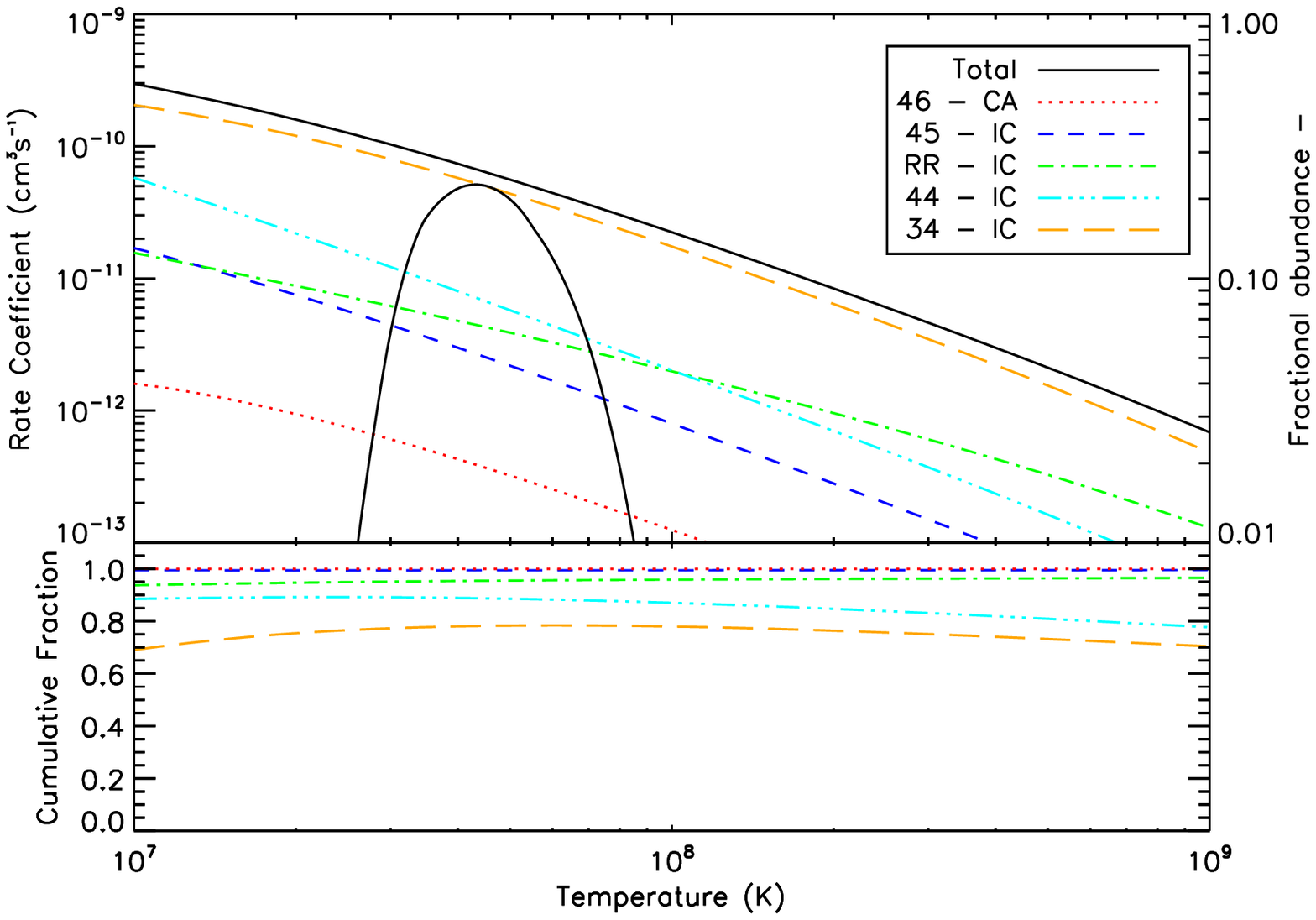}
\caption{DR rate coefficients for each core-excitation of 29-like tungsten, the RR rate coefficient, and the sum total of
these. The parabola is the ionization fraction for 29-like, calculated using P\"{u}tterich~\etal \protect\cite{putterich2008a}'s 
recombination rate coefficients, and Loch~\etal \protect\cite{loch2005a}'s ionization rate coefficients. The bottom plot gives the 
cumulative contribution of each recombination rate coeffiicent, and is described in text.}
\label{fig:culike}
\end{centering}
\end{figure}

\begin{figure}
\begin{centering}
\includegraphics[width=85mm]{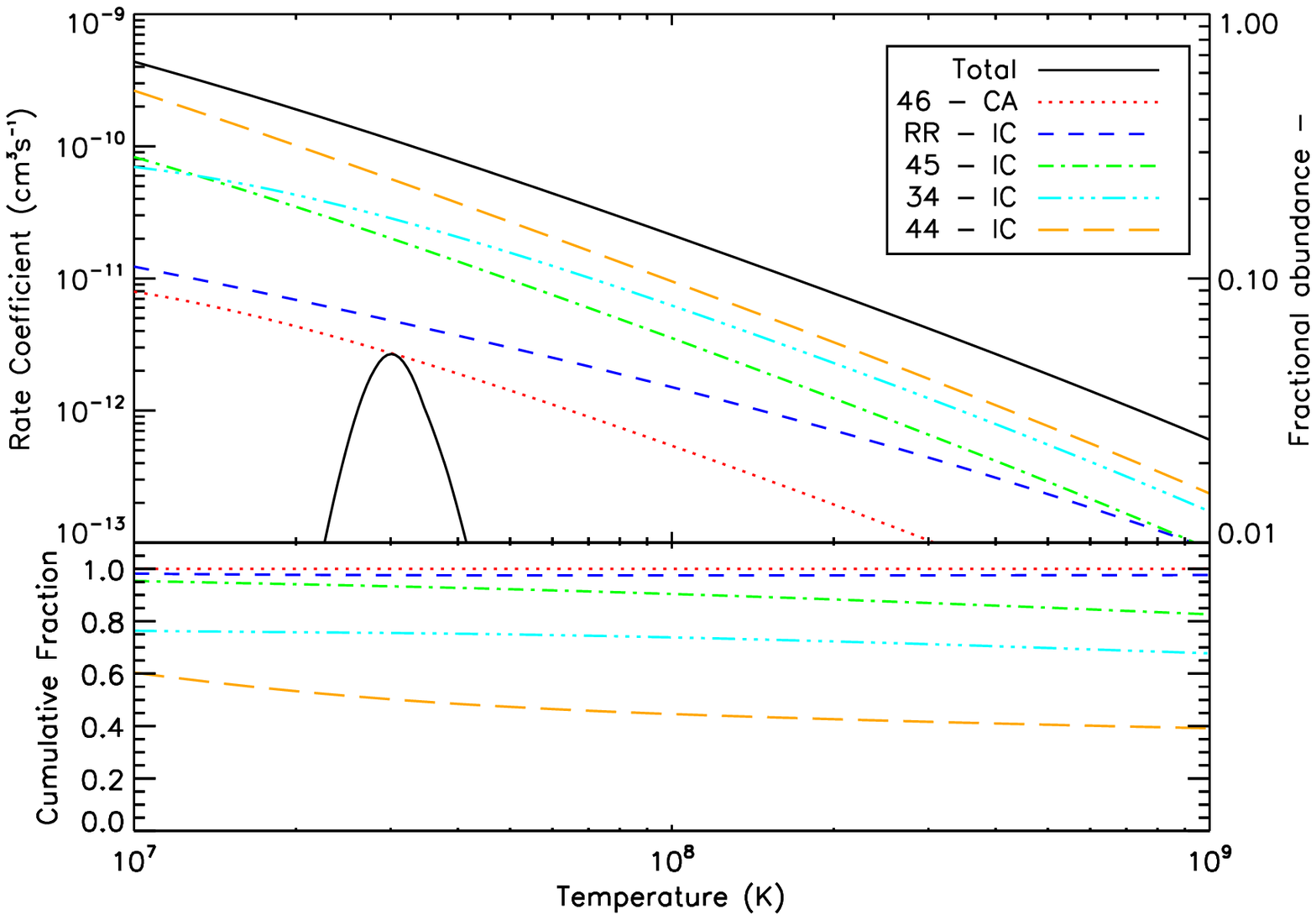}
\caption{The same as Figure \ref{fig:culike}, but for 34-like.}
\label{fig:selike}
\end{centering}
\end{figure}

The $4s^q$ ions cover 29-like and 30-like, and the $4s^{2}4p^{r}$ ions cover 31- 36-like. In Figures \ref{fig:culike} and
\ref{fig:selike} we have plotted the DR and RR rate coefficients for 29- and 34-like, along with the cumulative fraction 
described earlier.

\subsubsection{3--4}
\begin{figure}
\begin{centering}
\includegraphics[width=85mm]{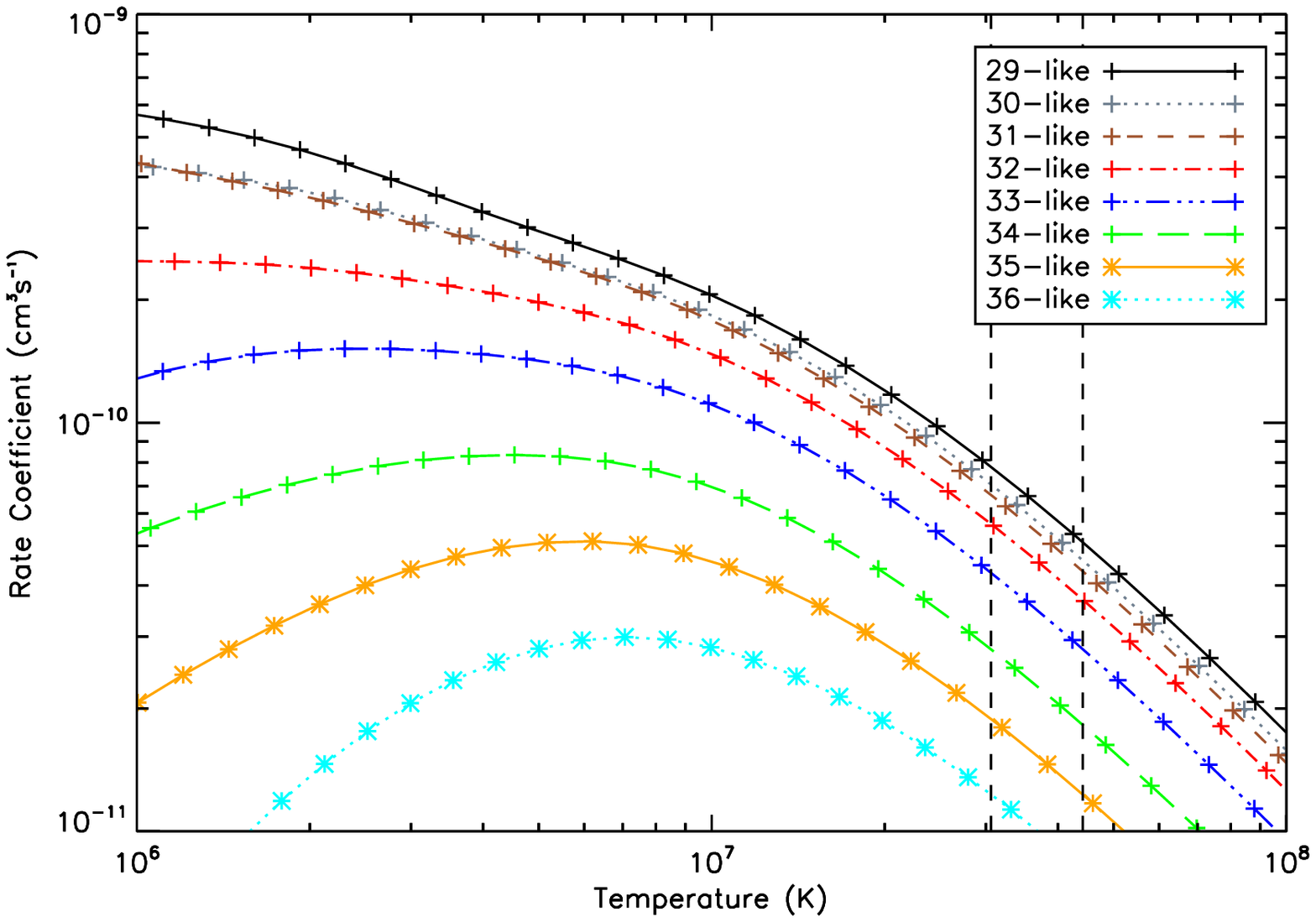}
\caption{Total DR rate coefficients for the 3--4 core-excitation for ionization stages 29- to 36-like tungsten calculated in IC.}
\label{fig:34rates2}
\end{centering}
\end{figure}

The 3--4 core-excitation is an inner-shell process, and remains a significant contributor to the total DR rate coefficient 
providing 78\% at peak abundance temperature for 29-like. This decreases as the $n=4$ shell is filled, contributing 
13\% by 36-like at peak abundance. The 3--4 core-excitation remains the dominant contributor to the total recombination rate coefficient 
until 32-like, giving way to 4--4 and 4--5 at ionization stages lower than this. In Figure \ref{fig:34rates2} we have plotted 
the 3--4 DR rate coefficients for 29- to 36-like calculated in IC, and have indicated the CP temperature region for these ions. 
As the $n=4$ shell fills, the number of possible promotions from $n=3$ decreases. This leads to a decrease in the DR rate 
coefficient.

\begin{figure}
\begin{centering}
\includegraphics[width=85mm]{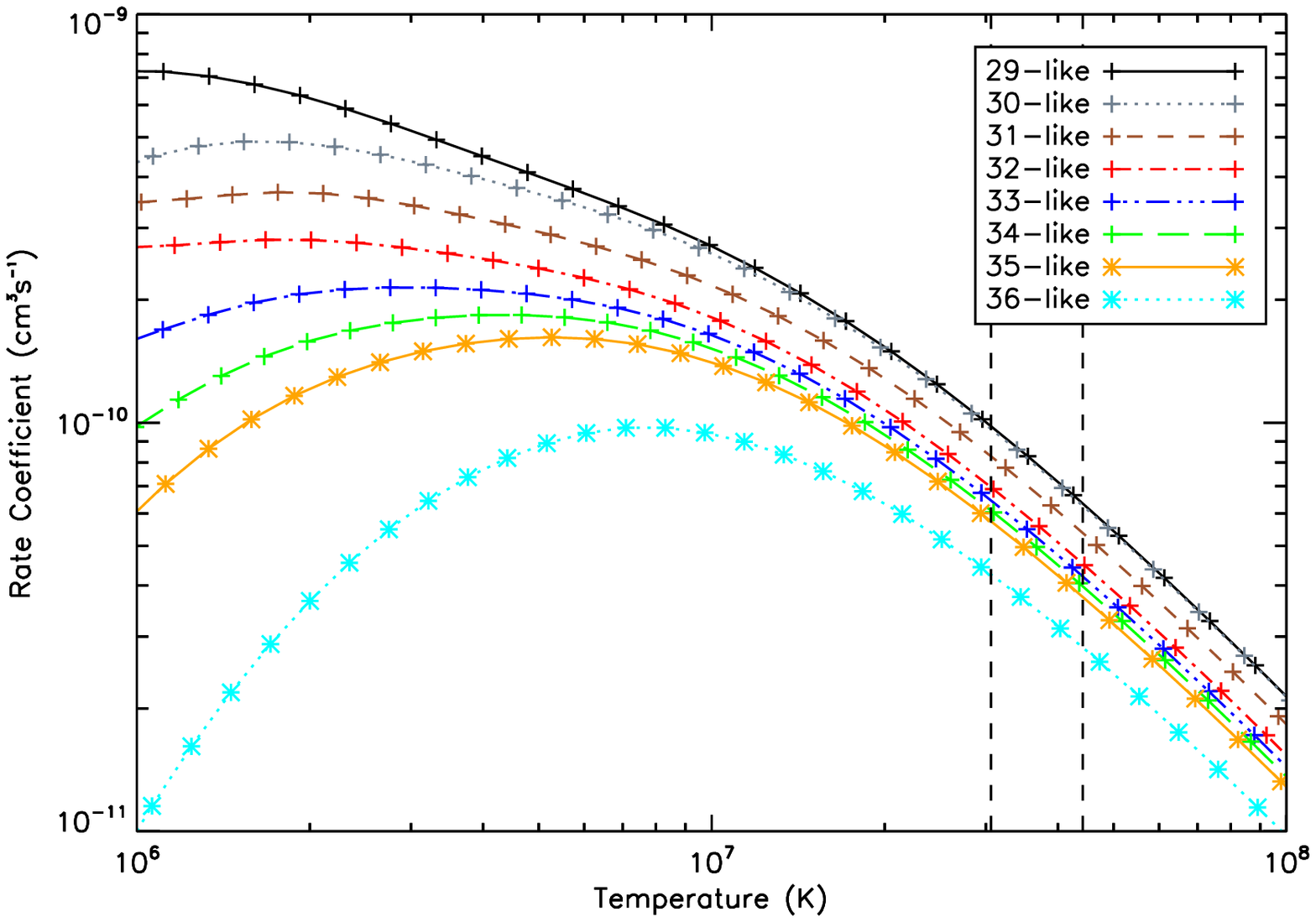}
\caption{The same as Figure \ref{fig:34rates2}, but calculated in CA.}
\label{fig:34rates2CA}
\end{centering}
\end{figure}

In Figure \ref{fig:34rates2CA} we have plotted the 3--4 DR rate coefficients for 29- to 36-like calculated in CA. By 
comparing Figures \ref{fig:34rates2} and \ref{fig:34rates2CA}, it can be seen that the total DR rate coefficient 
decreases much faster in the CP region in IC than it does for CA. The suppression effect also becomes much stronger with 
decreasing residual charge. For 29-like, the CA result is 22\% larger than the IC result at peak abundance, whereas 
for 36-like the CA result is 83\% larger than the IC result at peak abundance. This increase in the suppression effect 
is due to electrons being more tightly bound in the $3d$ shell as the $4s$ and $4p$ shells are filled, further increasing the 
Auger yield.

\subsubsection{4--4}
\begin{figure}
\begin{centering}
\includegraphics[width=85mm]{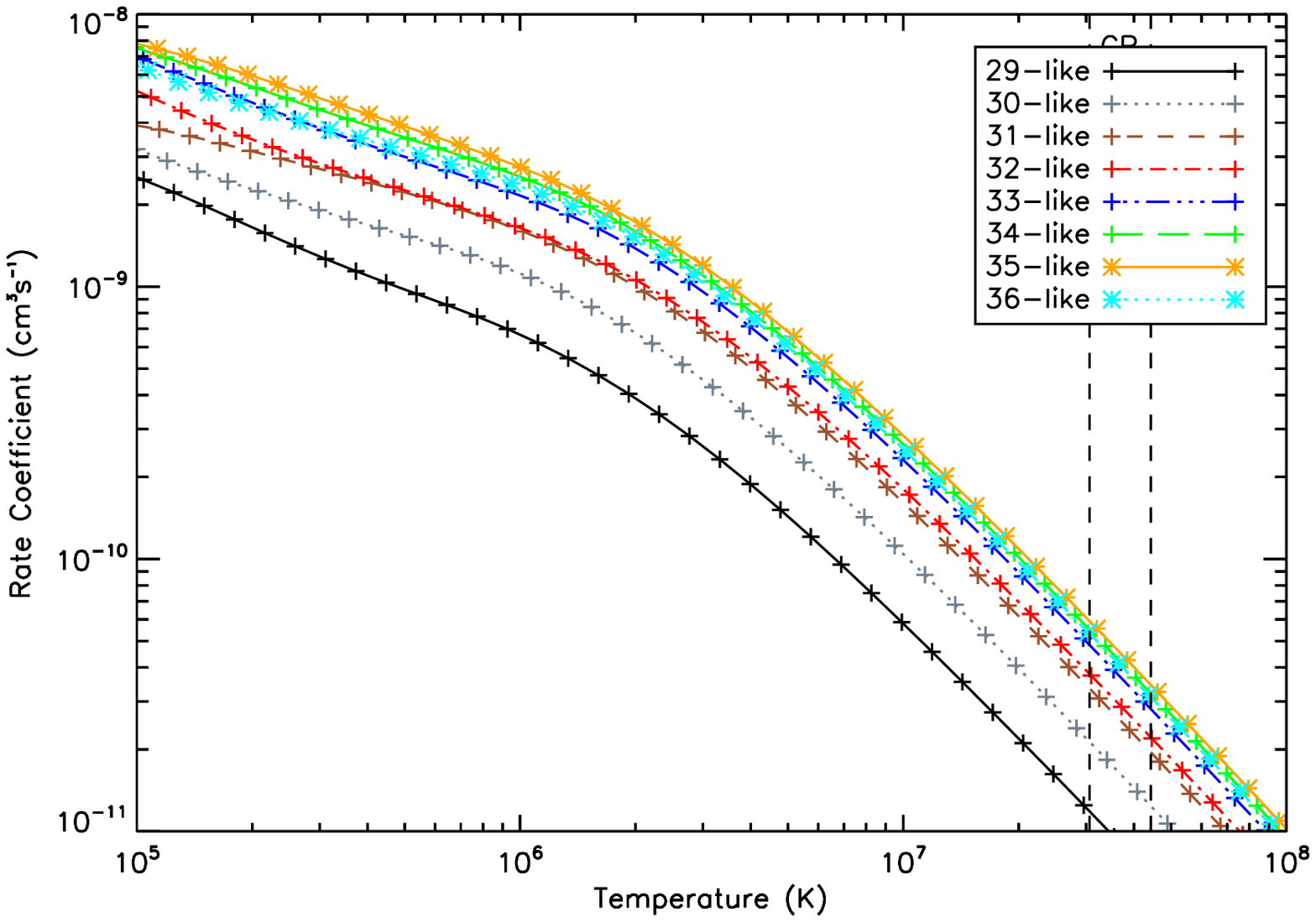}
\caption{Total DR rate coefficients for the 4--4 core-excitation for ionization stages 29- to 36-like tungsten calculated in IC.}
\label{fig:44ratesic}
\end{centering}
\end{figure}

The 4--4 core-excitation becomes the dominant contribution to the total recombination rate coefficient from 33-like onwards. 
This is a simple consequence of the $n=4$ shell filling and suppressing the 3--4 core-excitation. In Figure \ref{fig:44ratesic} 
we have plotted the 4--4 total DR rate coefficients for 29- to 36-like calculated in IC, and have also included the CP temperature 
region for these ions. For 29-like, 4--4 contributes 11\% to the total recombination rate coefficient at peak abundance, whereas for 36-like, 
4--4 contributes 58\%. The total DR rate coefficient jumps noticeably between 29- and 30-like. This is a simple consequence of adding a second 
electron which effectively doubles the DR rate coefficients of 29-like. A similar jump is seen between 30-like and 31-like. 
The DR rate coefficients are very similar for 31- and 32-like. The DR rate coefficients appear to jump again between 33- and 
36-like. In both the 31- to 32-like and 33- to 36-like the DR rate coefficients are confined to a tight band of values. This implies 
that 33- to 36-like are relatively insensitive to structural changes. With the exception of 36-like, the total DR rate 
coefficient appears to increase in tandem with a decreasing residual charge at peak abundance temperature. In the case of 
36-like, the total DR rate coefficient decreases sufficiently to place it between 33- and 34-like. This is due to the $4p$ 
shell being completely filled, and preventing $4p-4p$ transitions occuring.

\begin{figure}
\begin{centering}
\includegraphics[width=85mm]{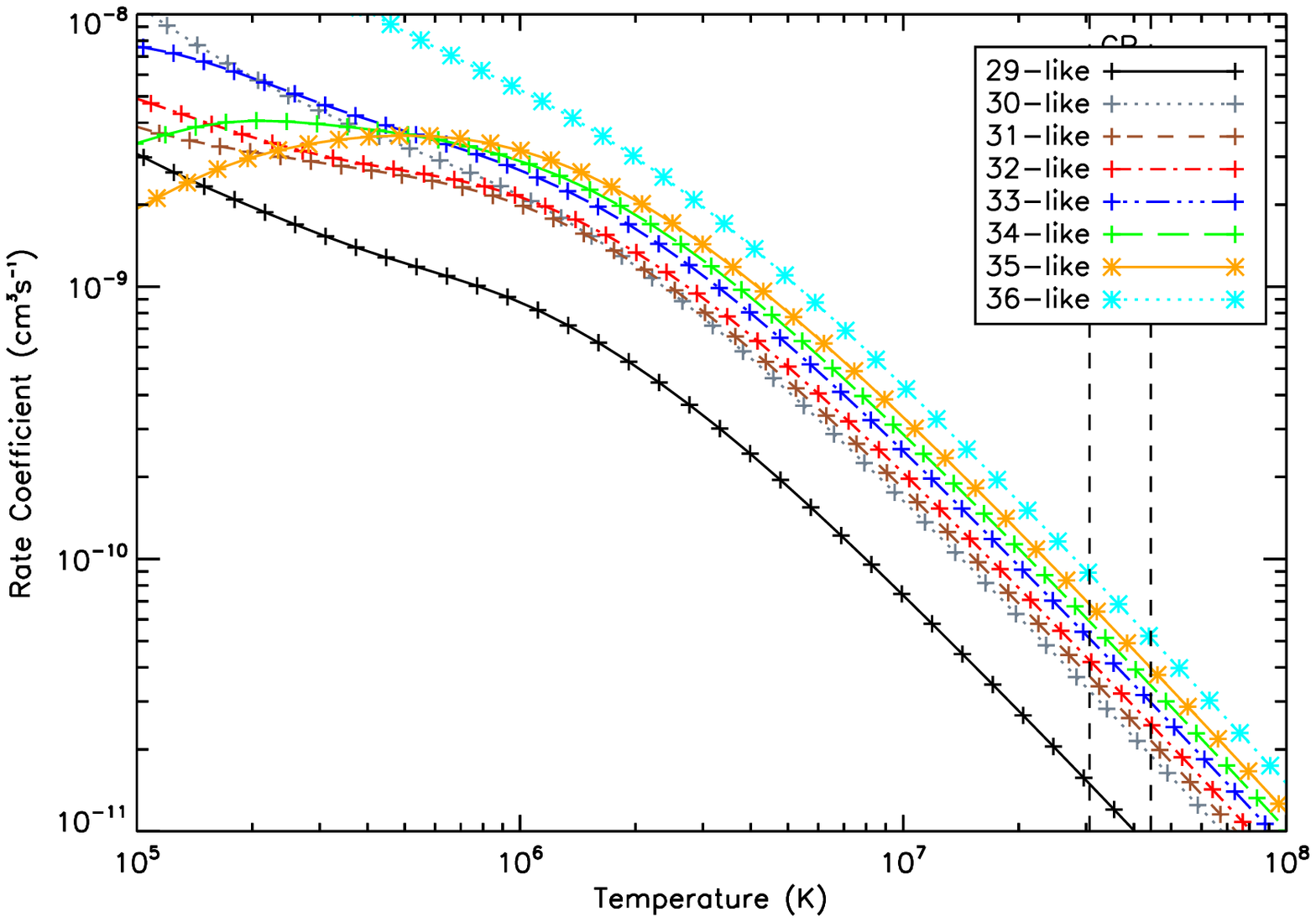}
\caption{The same as Figure \ref{fig:44ratesic}, but calculated in CA.}
\label{fig:44ratesca}
\end{centering}
\end{figure}

In Figure \ref{fig:44ratesca} we plot the DR rate coefficients again but calculated in CA. Similar to Figure \ref{fig:44ratesic} 
there is a jump in the total DR rate coefficient between 29- and 30-like, however, the banded structure observed in the IC 
results is not seen in the CA results. Instead, from 30-like the DR rate coefficient increases steadily with decreasing charge 
residual. It is also seen that, contrary to the IC results, the DR rate coefficients increase in tandem with a decreasing charge 
residual at peak abundance temperature, with 36-like being the largest. This is easily explained as the $4p-4p$ transition 
cannot occur in CA as it is an elastic transition.

\subsubsection{4--5}
\begin{figure}
\begin{centering}
\includegraphics[width=85mm]{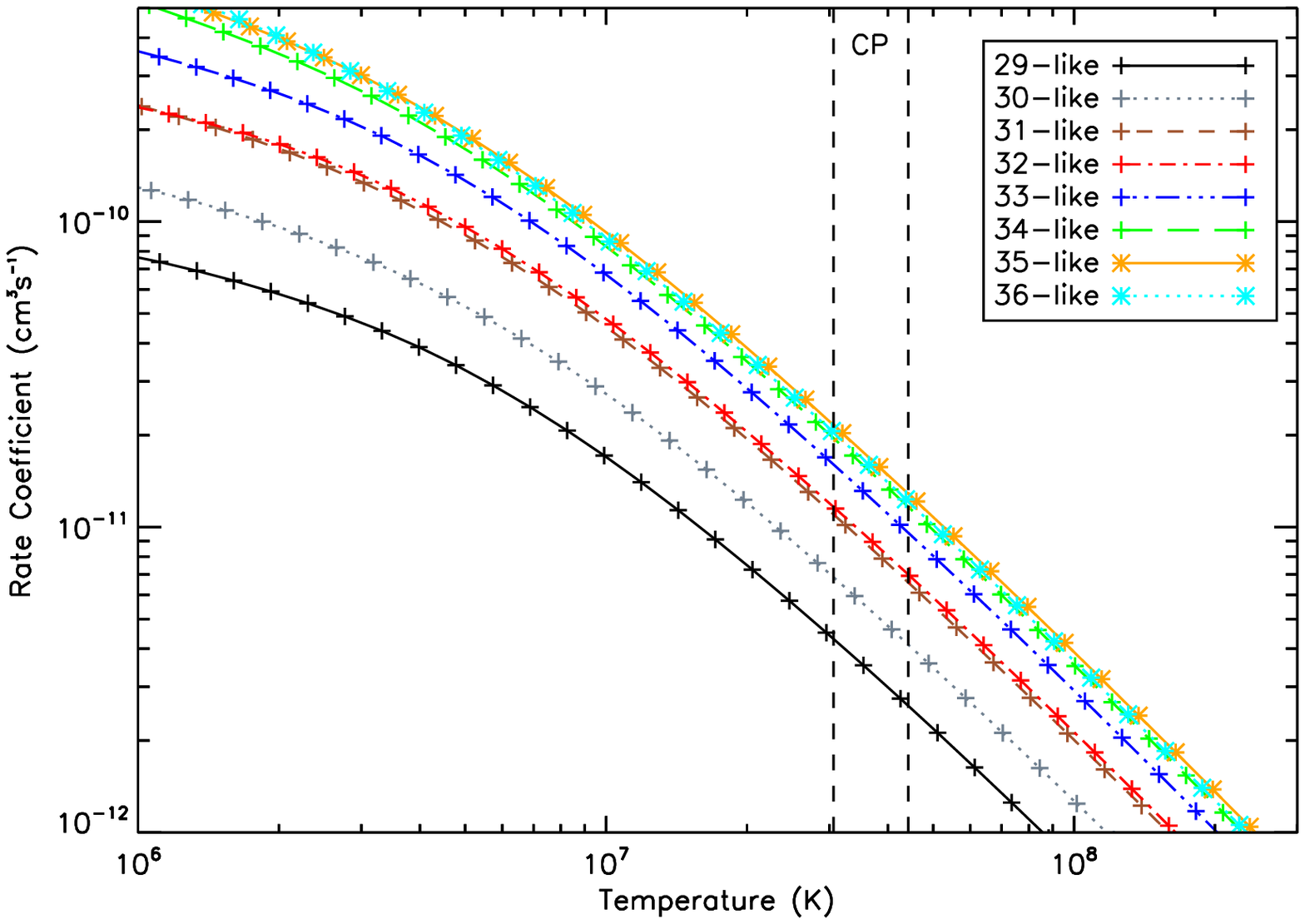}
\caption{Total DR rate coefficients for the 4--5 core-excitation for ionization stages 29- to 36-like tungsten calculated in IC.}
\label{fig:45ratesic}
\end{centering}
\end{figure}

The 4--5 core-excitation is initially comparatively small with respect to 3--4 and 4--4, increasing to be the second largest contribution to the 
recombination rate total by 36-like. For 29-like, 4--5 contributes 4\% at peak abundance. This steadily increases as we move along
the charge states, contributing 22\% at peak abundance by 36-like. In Figure \ref{fig:45ratesic} we have plotted the 4--5 DR rate 
coefficients for 29- to 36-like calculated in IC, and have also indicated the CP temperature region for these ions. As seen in the 4--4 core 
excitation, there appears to be a banded structure, where a large jump is seen between 29- and 30-like, and between 30- and 31-like.
Again, the 31- and 32-like DR rate coefficients are very similar. There is another large jump between 32- and 33-like. Interestingly,
in this case, 33-like is not grouped together with 34- to 36-like. Again, as observed in 4--4, the total DR rate coefficients for 36-like 

\begin{figure}
\begin{centering}
\includegraphics[width=85mm]{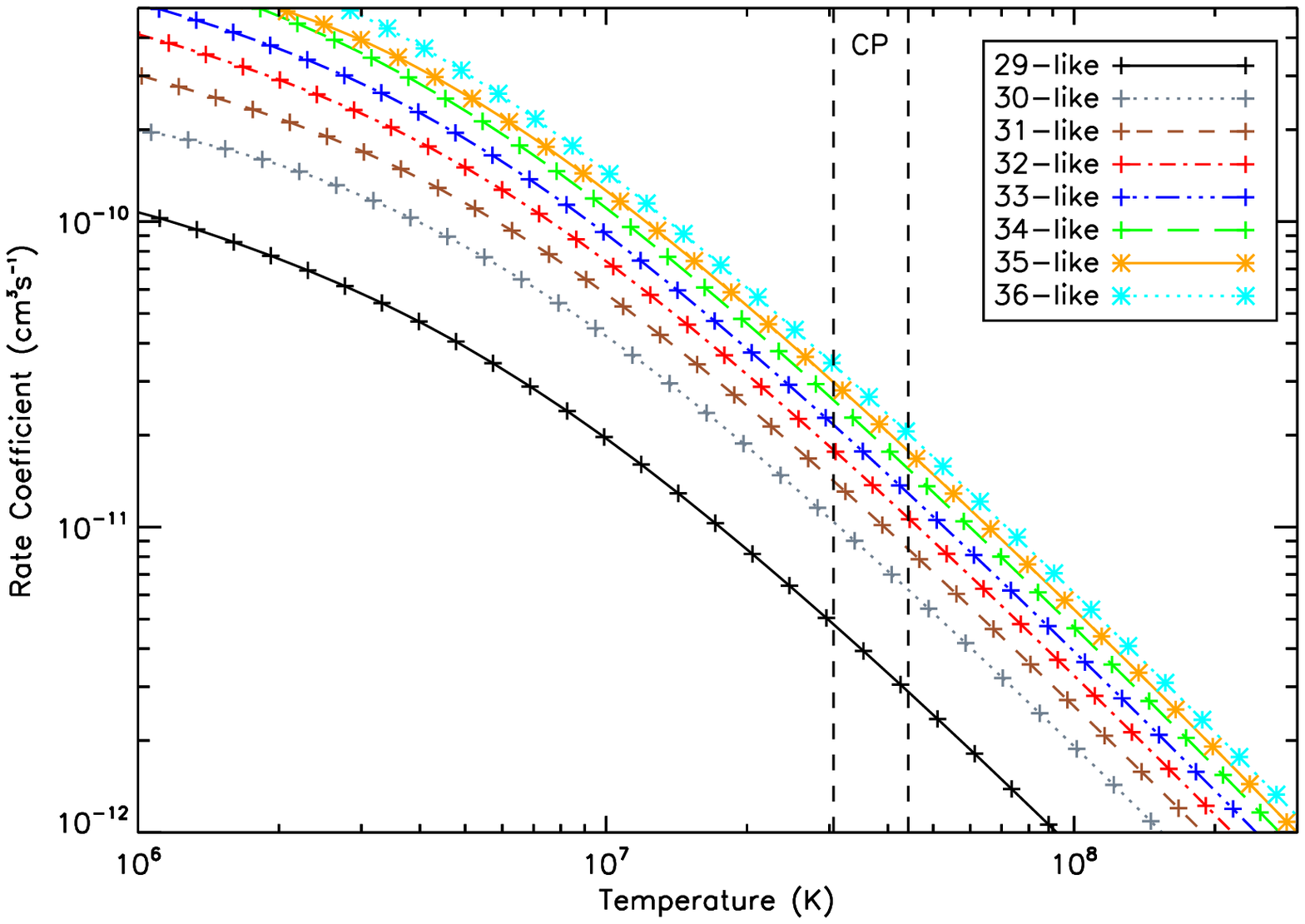}
\caption{The same as Figure \ref{fig:45ratesic}, but calculated in CA.}
\label{fig:45ratesca}
\end{centering}
\end{figure}

In Figure \ref{fig:45ratesca} we plot the total DR rate coefficients for 4--5 calculated in CA. 4--5 in CA bears many similarities to
4--4 in CA, in that only a single large jump in DR rate coefficient is seen between 29- and 30-like. In addition, from 30- to 
36-like, the DR rate coefficient increases in tandem with decreasing residual charge at peak abundance temperature with 36-like
being the largest.

\subsubsection{4--6}
\begin{figure}
\begin{centering}
\includegraphics[width=85mm]{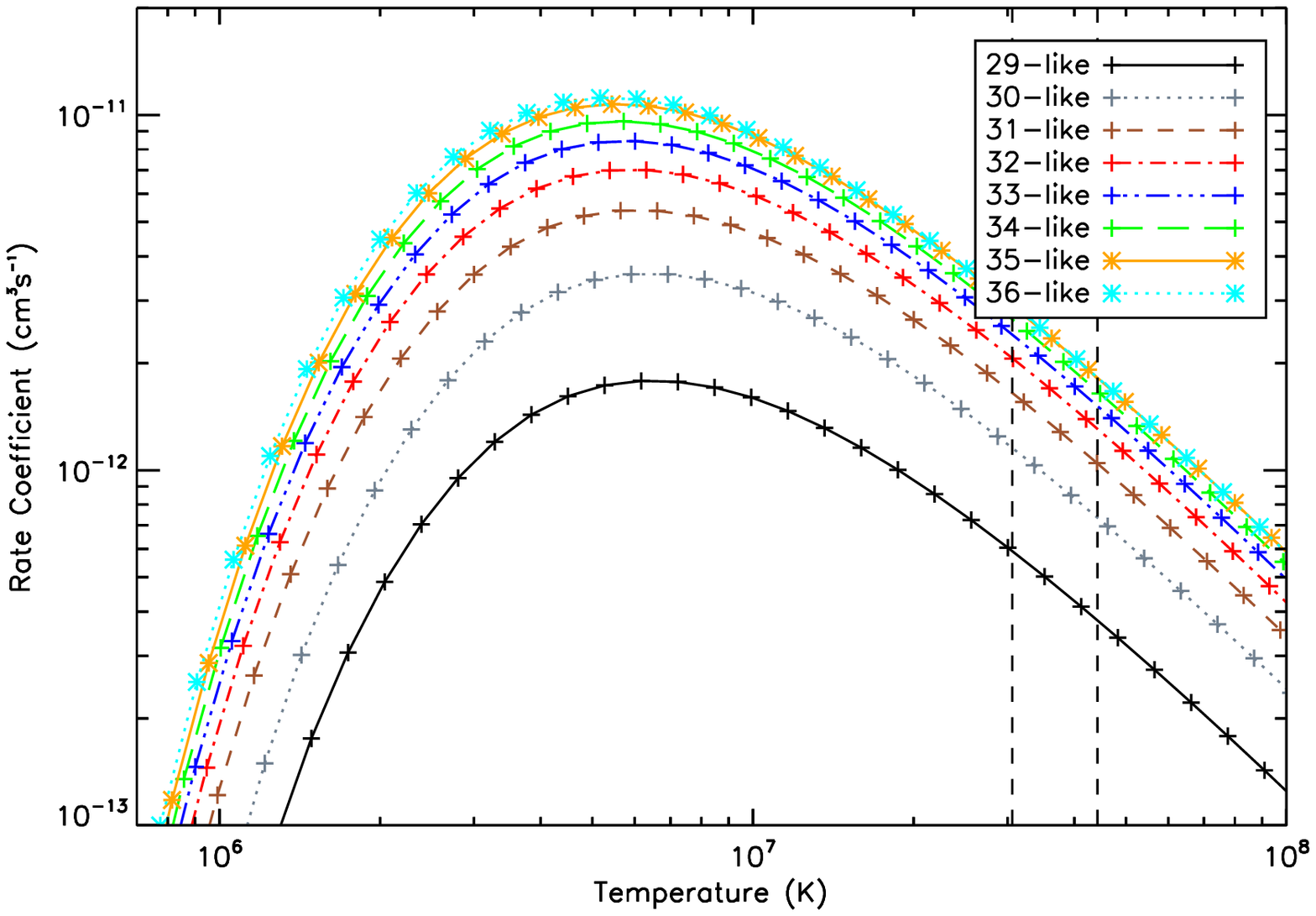}
\caption{Total DR rate coefficients for the 4--6 core-excitation for ionization stages 29- to 36-like tungsten calculated in CA.}
\label{fig:46rates}
\end{centering}
\end{figure}

The 4--6 core-excitation provides the smallest contribution to the total recombination rate coefficient. It has been included for 
completeness, and was calculated in CA. For 29-like, 4--6 contributes 0.6\% at peak abundance. This increases steadily
with decreasing residual charge, contributing 3\% by 36-like. In Figure \ref{fig:46rates} we have plotted the 4--6 DR 
rate coefficients for ionization stages 29- to 36-like. We have also indicated the CP temperature region for these ions. It can be seen that 
the rate coefficient spans $\sim{1}$ dex over 29- to 36-like. As the $n=4$ shell fills, more electrons can be excited to the $n=6$ 
shell, increasing the DR rate coefficient with decreasing residual charge.

\section{RR - Results}
RR becomes less important to the total recombination rate coefficient as the residual charge decreases. However, RR
is not negligible. For the ions considered, the RR rate coefficients are largest for 19-like, contributing 14\% to the total recombination rate
coefficient. This steadily decreases as the residual charge decreases. At peak abundance temperature, RR contributes 12\%,
9\%, and 8\% to the total recombination rate coefficient for 22-like, 25-like, and 28-like respectively for the M-Shell. In
the N-shell, RR contributes 7\% and 4\% to the total for 29-like and 36-like respectively at peak abundance temperature.

\section{Discussion - Relativistic Configuration Mixing}\label{relconsec}
\begin{figure}
\begin{centering}
\includegraphics[width=85mm]{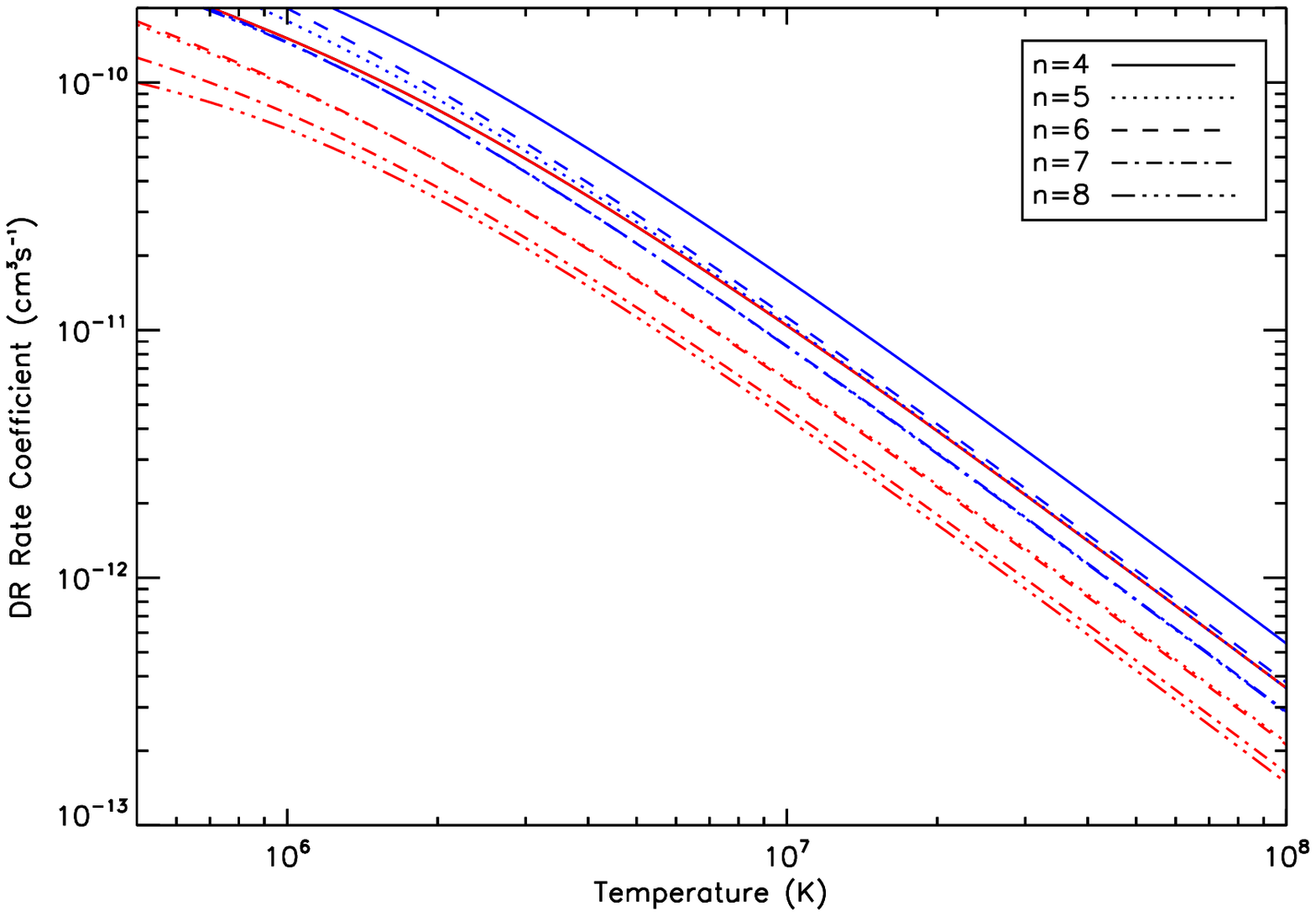}
\caption{Partial DR rate coefficients for 30-like 4--4 recombining into the $n=4-8$ complexes. The red curves indicate the
IC results, while the blue curves indicate the CA result.}
\label{fig:partials1}
\end{centering}
\end{figure}

\begin{figure}
\begin{centering}
\includegraphics[width=85mm]{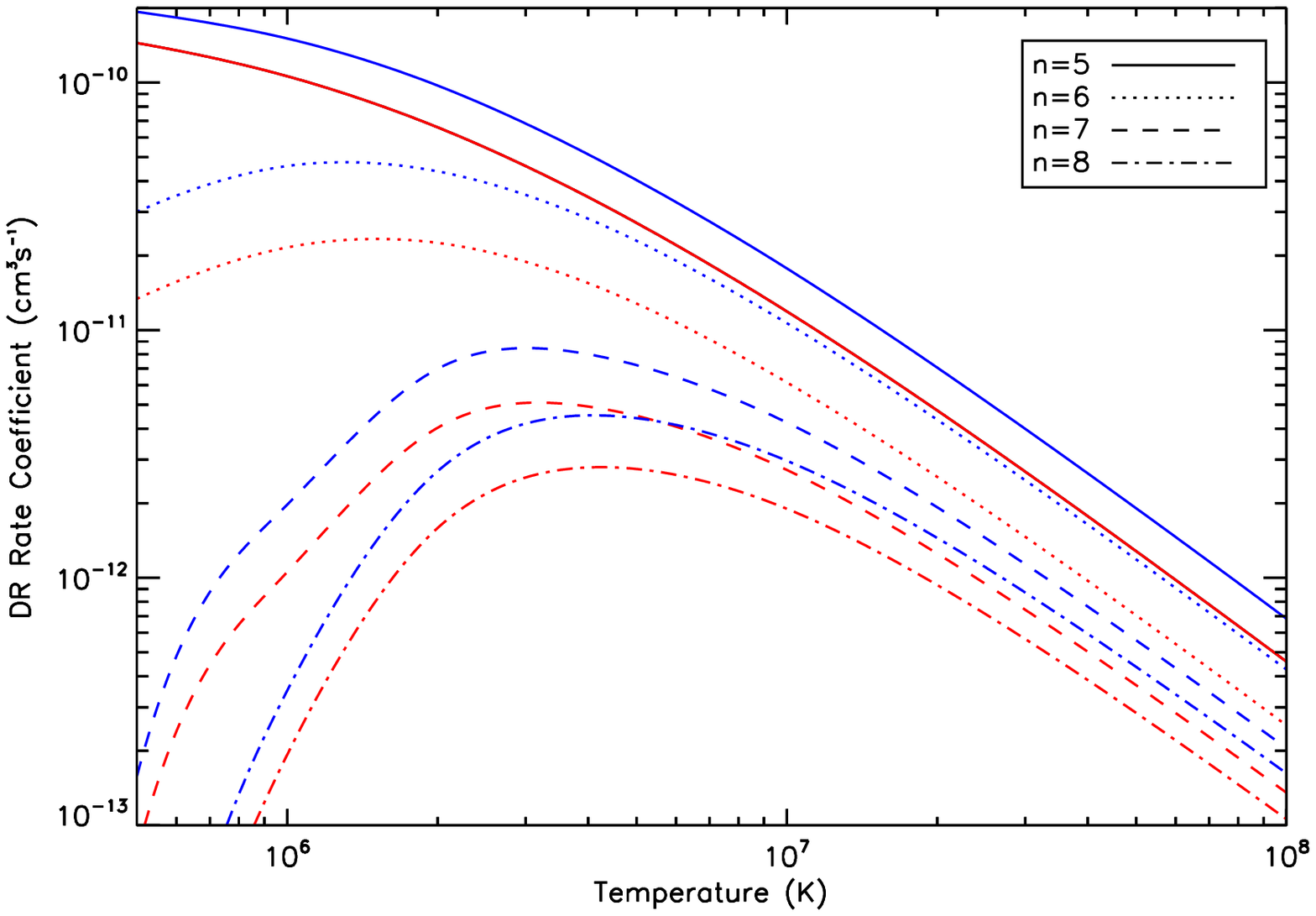}
\caption{Partial DR rate coefficients for 30-like 4--5 recombining into the $n=5-8$ complexes. The red curves indicate the
IC results, while the blue curves indicate the CA result.}
\label{fig:partials2}
\end{centering}
\end{figure}

As we move along the isonuclear sequence the atomic structure becomes more complex. With the addition of more electrons,
there are more possible ``one-up, one-down" configurations. The effect of mixing can be observed by comparing
results calculated in IC and CA. However, as mentioned in Preval~\etal \cite{preval2016a}, examining the differences between
IC and CA using only the totals can be misleading. Therefore, we must look to the partial DR rate coefficients. In
Figure \ref{fig:partials1} we have plotted the IC and CA partial DR rate coefficients for 30-like 4--4, where the
incident electron recombines into $n=4-8$. At peak abundance temperature, the $n=4-8$ CA partials differ from their IC
counterparts by 50\% to a factor $\sim{2}$. In Figure \ref{fig:partials2} we now plot the partial DR rate coefficients
for 30-like 4--5 recombining into $n=5-8$, again in IC and CA. At peak abundance temperature, the $n=5-8$ CA partials
differ from their IC counterparts by 50-70\%. Due to the large differences between IC and CA, these results highlight the 
importance of mixing, and also reinforces the need to use IC results when at all possible.

\section{Discussion - Comparison with other works}
There is little published data for the partial $3d^q$ DR rate coefficients, possibly because of the complexity of the calculations. 
However, as mentioned in the Introduction, Meng~\etal \cite{meng2009a} have calculated total DR rate coefficients for 27-like ($3d^9$) using
{\sc fac}. 28-like and beyond have multiple datasets available to compare with. Before comparing our results, we sum the DR rate coefficients for all core 
excitations for the charge-state concerned. Furthermore, we also omit the relativistic J\"{u}ttner correction for ease of comparison.
In all of our comparisons, the percentage difference between our DR rate coefficients $\alpha_{0}(T)$ and other authors' DR rate 
coefficients $\alpha_{i}(T)$ at some temperature $T$ is given as a fraction relative to our data, and is calculated as
\begin{equation}
\%(T) = 100\times\frac{\alpha_{i}(T) - \alpha_{0}(T)}{\alpha_{0}(T)}.
\end{equation}

\subsection{27-like}
The only data available for this ionization stage was calculated by Meng~\etal \cite{meng2009a}. In Table \ref{table:mengrates} we have tabulated 
the present DR rate coefficient for 27-like along with Meng~\etal's, and the \% difference between them. In general, Meng~\etal's
DR rate coefficients are larger than ours: at the peak abundance temperature ($5.97\times{10}^{7}$K), they differ 
from the current work by 31\%. The agreement between our and Meng~\etal's results (calculated using {\sc fac}) provides a benchmark against
{\sc autostructure}, and also shows that our calculation is reliable.

\subsection{28-like}
\begin{figure}
\begin{centering}
\includegraphics[width=85mm]{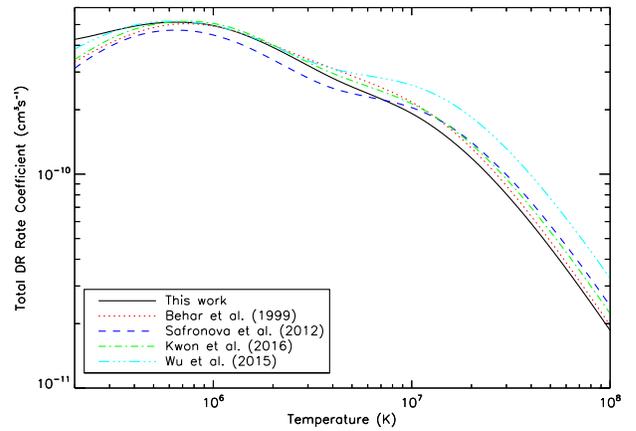}
\caption{Total DR rate coefficients of 28-like as calculated in this work (black solid), Behar~\etal
\protect\cite{behar1999b} (red dotted), Safronova~\etal \protect\cite{usafronova2012c} (blue dash), Kwon~\etal
\protect\cite{kwon2016b} (green dash-dot), and Wu~\etal \protect\cite{wu2015a} (cyan triple dot-dash).}
\label{fig:w46comp}
\end{centering}
\end{figure}

28-like is a closed shell ion ($3d^{10}$). It is a simple ion in terms of the possible DR channels available, and it is also 
important in plasma diagnostics, warranting much attention in the modelling community. In Figure \ref{fig:w46comp} we have plotted our total 
DR rate coefficients for 28-like, along with the results of Behar~\etal \cite{behar1999b}, Safronova~\etal \cite{usafronova2012b},
Kwon~\etal \cite{kwon2016a}, and Wu~\etal \cite{wu2015a}. We find that agreement between our DR rate coefficients, Behar~\etal's, 
and Kwon~\etal's is generally good at peak abundance temperature ($5.1\times{10}^{7}$K), differing by 7\% and 19\% 
respectively. However, agreement is much poorer for Safronova~\etal and Wu~\etal, differing from our rate coefficients by 27\% and 
70\%, respectively. In all cases, there are large discrepancies exceeding 20\% between our data and that of the other four authors at 
low temperature ($\sim{1}\times{10}^{5}$K). However, these variations occur well outside relevant plasma temperatures. Such variations are likely caused by the 
positioning of low temperature DR resonances, which are present in our calculation. The good agreement between Behar~\etal, Kwon~\etal, and our results
show that the calculations are reliable. In the case of the outliers, this gives an idea of the uncertainty in the calculation of
the total DR rate coefficients.

\subsection{29-like}
\begin{figure}
\begin{centering}
\includegraphics[width=85mm]{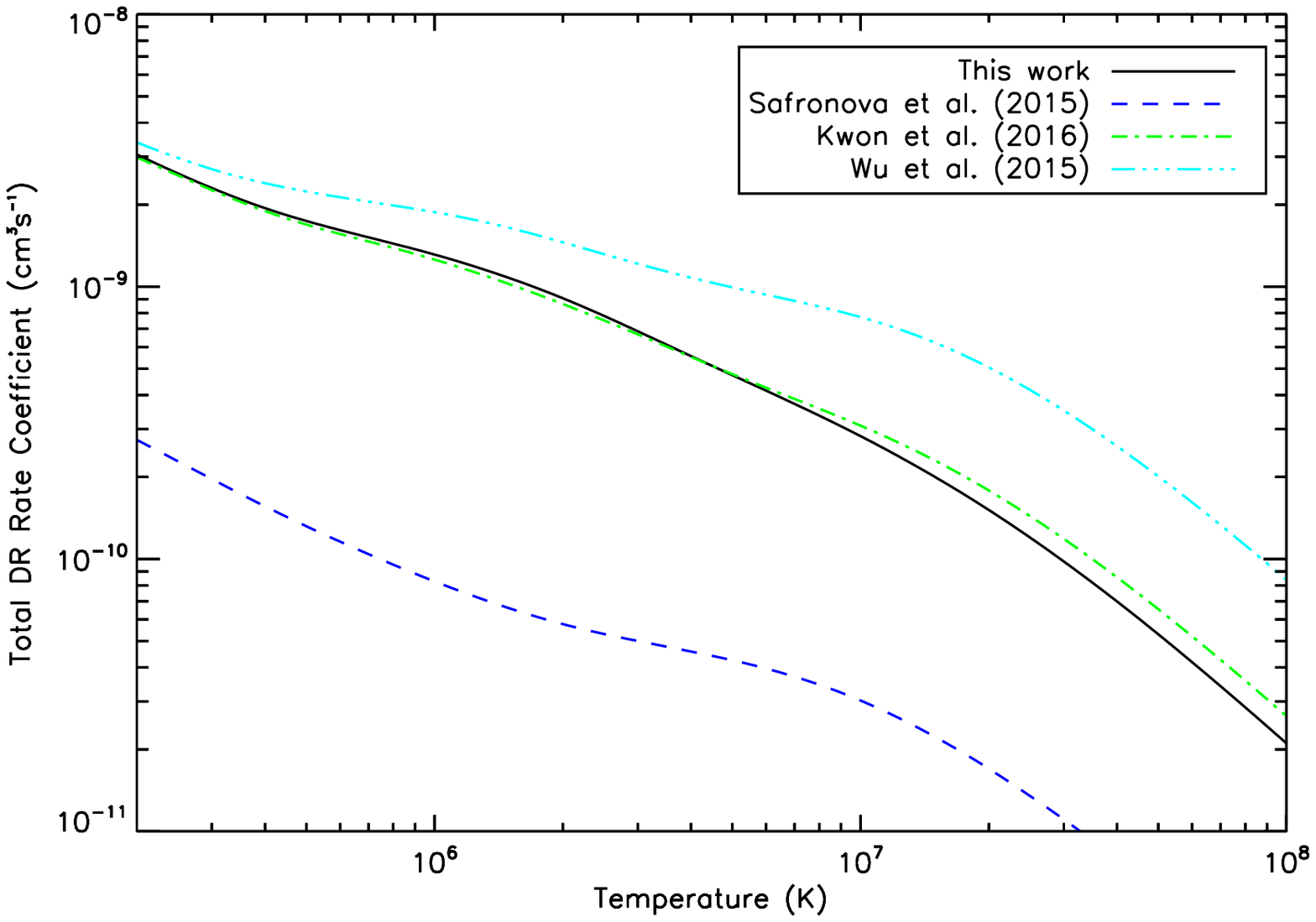}
\caption{Total DR rate coefficients of 29-like tungsten as calculated in this work (black solid),
Safronova~\etal \protect\cite{usafronova2015a} (blue dash), Kwon~\etal \protect\cite{kwon2016b} (green dash-dot),
and Wu~\etal \protect\cite{wu2015a} (cyan triple dot-dash).}
\label{fig:w45comp}
\end{centering}
\end{figure}

In Figure \ref{fig:w45comp} we have plotted the total DR rate coefficients for 29-like from this work, Safronova~\etal \cite{usafronova2015a} (using {\sc hullac}),
Kwon~\etal \cite{kwon2016a}, and Wu~\etal \cite{wu2015a} (both using {\sc fac}). Prior to the present work, there was no consensus between calculations. 
Good agreement is seen between our data and Kwon's at peak abundance 
temperature ($4.32\times{10}^{7}$K) differing by 23\%. Interestingly, we see even better agreement between our 
rate coefficients and Kwon's at lower temperatures ($\sim{1}\times{10}^{5}$K), with the difference being 2\% in places. However, from temperatures 
exceeding $4.5\times{10}^{6}$K, the difference gradually increases, reaching 26\% by $1\times{10}^{8}$K, where the rate coefficient is falling off rapidly.

Large differences between our DR rate coefficients and Wu~\etal's and Safronova~\etal's are observed over all temperature ranges.
In the case of Safronova~\etal, their results are a factor 11 smaller than our rate coefficients at low temperatures ($\sim{1}\times{10}^{5}$K).
With increasing temperature, this difference increases to a factor 16. Towards higher temperatures where $T^{-3/2}$ behaviour dominates, the difference 
stabilises to a factor 9. The size and consistency of this discrepancy implies that there is a systematic uncertainty in
Safronova~\etal's rate coefficients which is difficult to account for. 

We note that discrepancies of the size observed are not limited to this particular ion. For example, in our previous work 
\cite{preval2016a}, we noted a discrepancy of 47\% between our total DR rate coefficients and that of Safronova~\etal 
\cite{usafronova2009b} for 10-like tungsten at high temperatures, but good agreement with the calculations of 
Behar~\etal \cite{behar1999a}, differing from our results by $<10$\%.

In the case of Wu~\etal, even larger discrepancies are seen. While at low temperatures the agreement is better than $10$\%, the difference 
between our and Wu's data increases gradually to a factor $\sim{4}$ at the highest temperatures compared. The difference at peak abundance 
temperature is a factor ${3.7}$. We defer discussion of this discrepancy to Section \ref{sec:aug}.

\subsection{30-like}
\begin{figure}
\begin{centering}
\includegraphics[width=85mm]{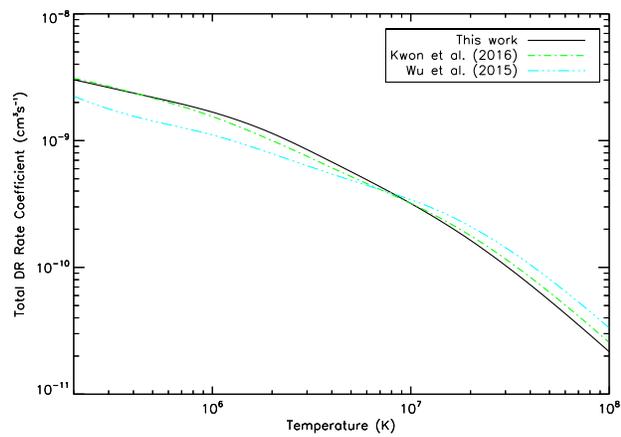}
\caption{Total DR rate coefficients of 30-like tungsten as calculated in this work (solid line),
Kwon~\etal \protect\cite{kwon2016b} (green dash-dot), and Wu~\etal \protect\cite{wu2015a} (cyan triple dot-dash).}
\label{fig:w44comp}
\end{centering}
\end{figure}

In Figure \ref{fig:w44comp} we have plotted the total DR rate coefficients for 30-like from this work, Kwon~\etal \cite{kwon2016a}, and 
Wu~\etal \cite{wu2015a}. As with 29-like, good agreement is seen between our total DR rate coefficients and Kwon~\etal's over the entire 
temperature range considered. At peak abundance temperature ($3.7\times{10}^{7}$K) Kwon~\etal's DR rate coefficients 
differ from ours by 14\%. Excellent agreement is seen towards lower temperatures, decreasing steadily to 0.5\% by $\sim{1\times{10}^{5}}$K.
Towards higher temperatures, the agreement deteriorates slightly, reaching 18\% by $\sim{1\times{10}^{8}}$K.

In the case of Wu~\etal's results, agreement is markedly better for 30-like than it was for 29-like. At peak abundance temperature the difference between
our and Wu~\etal's DR rate coefficients is 43\%. Towards lower temperatures this difference increases to 49\% before decreasing to 14\%
at $\sim{1\times{10}^{5}}$K. Towards higher temperatures, the agreement deteriorates further, reaching 55\% by $\sim{1\times{10}^{8}}$K.

\subsection{31- 36-like}
\begin{figure}
\begin{centering}
\includegraphics[width=85mm]{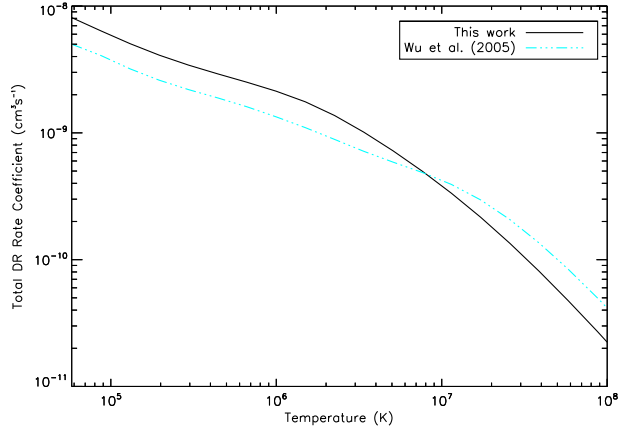}
\caption{Total DR rate coefficients of 32-like tungsten as calculated in this work (black solid),
and Wu~\etal \protect\cite{wu2015a} (cyan triple dot-dash).}
\label{fig:wu32comp}
\end{centering}
\end{figure}

In Table \ref{table:wutable} we list the \% difference between our and Wu~\etal's total DR rate coefficients over a range of 
temperatures. Agreement between our and Wu~\etal's varies, however, agreement appears to be worse generally at the extremes of
low and high temperature. In the case of 31-like, there is generally good agreement at low temperature, where our and Wu~\etal's data differ by 10\%
between $1.2\times{10}^{5}$ and $4.2\times{10}^{6}$K. Beyond $2.7\times{10}^{6}$K, our DR rate coefficients gradually become
smaller, differing by 88\% at peak abundance temperature from Wu~\etal. Similar differences occur for the following charge states. In Figure
\ref{fig:wu32comp} we plot an example of this for 32-like. At peak abundance temperature, our and Wu~\etal's total DR rate coefficients differ
by factors of 1.6--2.4 for 32-, 33-, 35-, and 36-like.

\begin{figure}
\begin{centering}
\includegraphics[width=85mm]{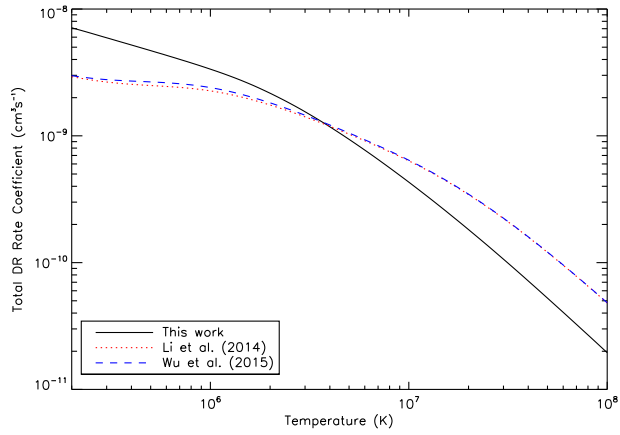}
\caption{Total DR rate coefficients of 35-like tungsten as calculated in this work (black solid),
Li~\etal \protect\cite{li2014a} (red dot), and Wu~\etal \protect\cite{wu2015a} (blue dash).}
\label{fig:w39comp}
\end{centering}
\end{figure}

In addition to the Wu~\etal data for 35-like, Li~\etal has also calculated DR rate coefficients for this ionization stage. In 
Figure \ref{fig:w39comp} we have plotted our 35-like total DR rate coefficients, and the corresponding DR rate coefficients of 
Li~\etal and Wu~\etal. It can be immediately seen that the size and shape of Li~\etal's and Wu~\etal's total DR rate coefficients are 
strikingly similar, differing at most by 7\% at $5.6\times{10}^{5}$K implying a similar methodology in their {\sc fac}
calculations. Towards lower temperatures our DR rate coefficients are larger than Li's differing by 61\% at 
$\sim{1\times{10}^{5}}$K. Towards higher temperatures the differences are much larger, differing by a factor $\sim{2.5}$ at $\sim{1\times{10}^{8}}$K. At peak abundance 
temperature ($2.9\times{10}^{7}$K) the difference between our data and Li's is a factor $\sim{2}$.

\subsection{Auger Supression}\label{sec:aug}
\begin{figure}
\begin{centering}
\includegraphics[width=85mm]{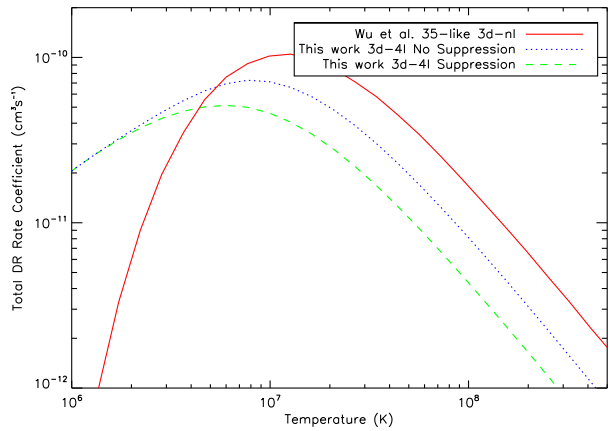}
\caption{Wu~\etal's \protect\cite{wu2015a} DR rate coefficients for 35-like tungsten $3d-nl$ (red solid) compared
to our 35-like $3d-4l$ calculations excluding and including suppression due to core-rearrangement (blue dotted and green dashed respectively).}
\label{fig:coresupp}
\end{centering}
\end{figure}

When comparing our total DR rate coefficients with Wu~\etal for 29-like to 36-like, we saw that Wu~\etal's rate coefficients 
were consistently larger than ours by significant amounts at low and high temperatures. Kwon~\etal \cite{kwon2016b} noted 
this difference, and attibuted the low temperature differences to Wu~\etal using a simplified set of configurations in their calculations. We do not 
discuss the low temperature regime further. At high temperatures, Auger suppression 
effects arising from core rearrangement for inner-shell DR become important. The inclusion of these core-rearrangement configurations is computationally demanding.
If core-rearrangement is neglected in inner-shell calculations, the resultant DR rate will be artificially inflated. For 35-like, Wu~\etal has plotted the $3s-$, 
$3p-$, and $3d-nl$ contributions to the total DR rate coefficient (cf. figure 6 in \cite{wu2015a}). In an attempt to explain the 
discrepancy between our and Wu~\etal's results, we repeated our 3--4 core-excitation calculation with the same structure as before, 
but neglect core-rearrangement. This also partially removes the contribution from radiation into autoionizing states. As our 
3--4 calculation includes only excitations from $3d$, we only compare Wu's $3d$ contribution. In Figure \ref{fig:coresupp} we 
plot our $3d-n\ell$ DR rate coefficients calculated with and without core rearrangement and compare them to Wu's $3d-n\ell$ calculations. It can be seen that our 
non-suppressed calculation is much closer to the Wu result than the suppressed calculation. At peak abundance temperature, the 
DR rate coefficients including core rearrangement differs from Wu~\etal's by a factor $\sim{3}$, whereas excluding the core 
rearrangement decreases this difference to 87\%. The remaining difference may come from other core-excitations such as 3--5, 3--6, and so on. 

\section{Discussion - Comparison with P\"{u}tterich~\etal}
\begin{figure}
\begin{centering}
\includegraphics[width=85mm]{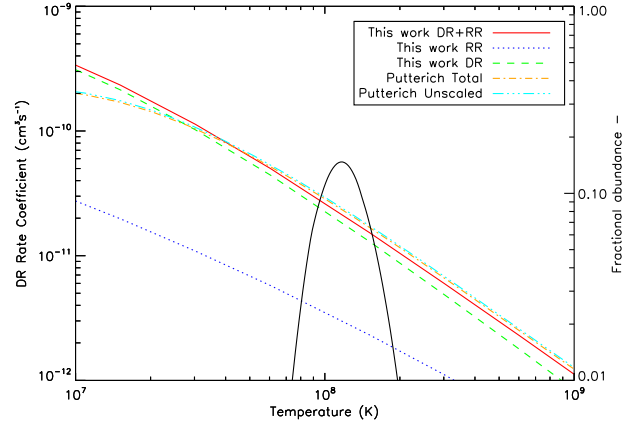}
\caption{Total recombination rate coefficients for 19-like from this work (solid red), and the scaled (orange
dash-dot) and unscaled (cyan triple dot-dash) results from P\"{u}tterich~\etal \cite{putterich2008a}. We also include the DR and RR contributions
that compose our total recombination rate coefficient. The black parabola is the ionization fraction for this ionization state
calculated using the recombination rate coefficients of P\"{u}tterich~\etal \cite{putterich2008a}, and the ionization rate
coefficients of Loch~\etal \protect\cite{loch2005a}.}
\label{fig:19likecomp}
\end{centering}
\end{figure}

Agreement between our recombination rate totals and P\"{u}tterich~\etal's \cite{putterich2008a} is
very good from 19- to 26-like. Recall that P\"{u}tterich~\etal scaled their recombination rate coefficients by empirically
determined scaling factors to improve agreement with observed spectral emission. In Figure \ref{fig:19likecomp} we have plotted
the present recombination rate coefficients and P\"{u}tterich~\etal's scaled and unscaled results for 19-like. We have also
included our individual DR and RR contributions to the total. At peak abundance temperature the present recombination
rate coefficients for 19-like differ from P\"{u}tterich~\etal's scaled data by 9\%. The unscaled data for 19-like bears a
similar level of agreement with our results, differing by 12\% at peak abundance temperature. Similarly, for 24-
and 26-like, the differences are 29\% and 13\% respectively for the scaled data. For the unscaled data,
the differences are 31\% and 45\% respectively.

\begin{figure}
\begin{centering}
\includegraphics[width=85mm]{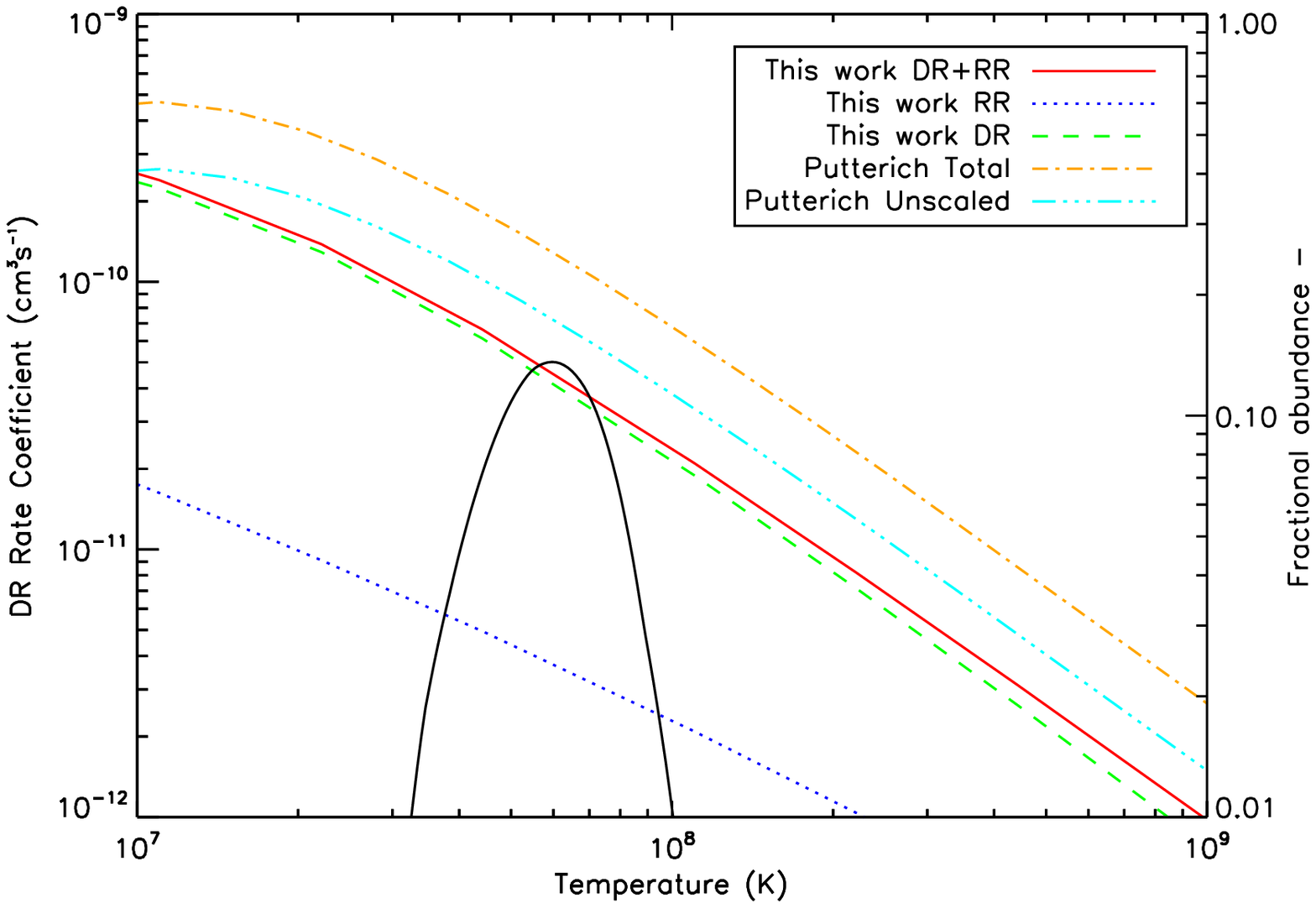}
\caption{The same as Figure \protect\ref{fig:19likecomp}, but for 27-like.}
\label{fig:27likecomp}
\end{centering}
\end{figure}

Beyond 26-like, the agreement between ours and P\"{u}tterich~\etal's scaled data deteriorates sharply. In Figure
\ref{fig:27likecomp} we have plotted the present recombination rate coefficients and P\"{u}tterich~\etal's scaled and
unscaled results for 27-like, again including the present DR and RR contributions. At peak abundance temperature,
our and P\"{u}tterich~\etal's scaled data differ by a factor $\sim{3}$. Interestingly, the unscaled data is in better agreement with
our results, where the difference is 57\% at peak abundance temperature. For 28-like, the scaled data agrees
better than the unscaled data, differing with our results by 29\% and 81\% respectively. From 32-like
onwards the unscaled data agrees better with our results than the scaled data. However, it should be stressed that while
agreement is improved, the difference between our results and the unscaled data is still very large, being 49\%
for 32-like, increasing to a factor $\sim{2}$ for 36-like.

\section{Discussion - Zero density Ionization balance}\label{ionbal}
Using our recombination rate coefficients, we calculate the ionization balance to quantify any changes that occur. To 
calculate the ionization fraction, we replace the relevant recombination rate coefficients from P\"{u}tterich~\etal \cite{putterich2008a} 
with our data. We use this recombination data in conjunction with the ionization rate coefficients of Loch~\etal \cite{loch2005a}. 

\begin{figure}
\begin{centering}
\includegraphics[width=85mm]{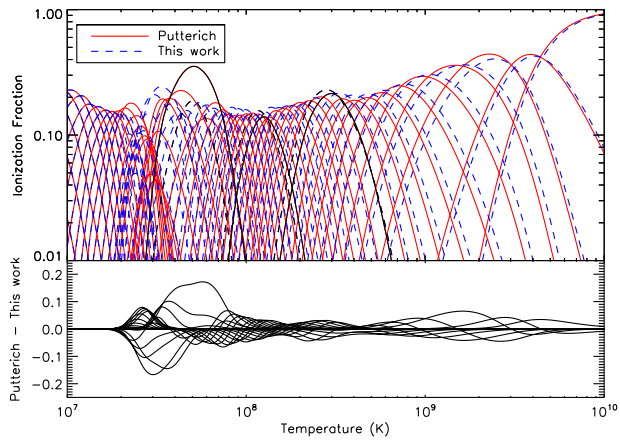}
\caption{Tungsten ionization fraction using P\"{u}tterich~\etal's \cite{putterich2008a} recombination rate coefficients (solid) and
our recombination rate coefficients (dashed) up to 36-like. Both ionization fractions use the ionization rate coefficients of Loch~\etal 
\protect\cite{loch2005a}. From left to right, we have marked the fractions of 28-like, 18-like, and 10-like in 
black. The bottom plot shows the difference between P\"{u}tterich~\etal's and our ionization fraction.}
\label{fig:ionbalnew}
\end{centering}
\end{figure}

\begin{figure}
\begin{centering}
\includegraphics[width=85mm]{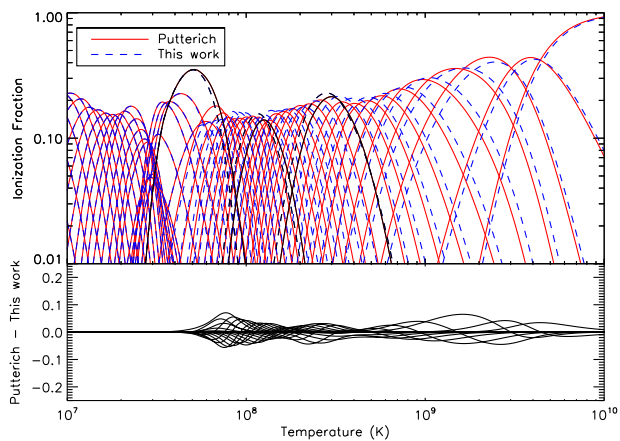}
\caption{The same as Figure \protect\ref{fig:ionbalnew}, but including our recombination data up to 26-like only.}
\label{fig:ionbalres}
\end{centering}
\end{figure}

In Figure \ref{fig:ionbalnew} we have plotted the ionization fraction using only the P\"{u}tterich~\etal recombination data, and the ionization fraction 
including our data. We have also plotted the difference between the two fractions, which is calculated by subtracting
our ionization fraction from P\"{u}tterich~\etal's for each charge state. Both ionization fractions use Loch's ionization 
rate coefficient data. Note that we have removed the J\"{u}ttner relativistic
correction from our recombination rate coefficients to simplify comparisons. It can be seen that the two ionization fractions are 
quite similar up to 27-like. However, for 28-like, we see a significant difference in the position and height of the ionization
fraction. This is caused by the large differences in our and P\"{u}tterich~\etal's recombination rate coefficients discussed 
in the previous section. To show this, we have replotted the ionization fraction in Figure \ref{fig:ionbalres}, but include 
our data only up to 26-like. It can now be seen that the large difference between the ionization fractions has now disappeared.

Including our new data up to 36-like has significantly altered the peak abundances and the peak abundance temperatures. In 
Table \ref{table:peaktemp} we have listed the peak abundances and temperatures as calculated when using P\"{u}tterich~\etal's and our 
ionization fractions for 01-like to 36-like. Most notable are the changes to the peak abundance for 28-like, which as mentioned 
previously, is an important ion in plasma diagnostics. While the peak temperature only changes by $<2$\% from P\"{u}tterich~\etal 
to the current work, the fractional abundance drops by 88\% from 0.35 to 0.19. The largest peak temperature change 
occurs for 25-like, decreasing by 13\% from $7.4\times{10}^{7}$ to $6.6\times{10}^{7}$K. Overall, there does not
appear to be a particular trend dictating whether a temperature or fraction will increase or decrease. Furthermore, these 
temperatures and fractions have been determined in the zero density approximation. The true effect of these changes will only
be seen with CR modelling.

\section{Conclusions}\label{concludesec}
We have compared our DR rate coefficients with multiple authors. Multiple datasets were available for 28-like, which is an
important ion in plasma diagnostics. At peak abundance temperature, our DR rate coefficients agree well with Behar~\etal \cite{behar1999b}
and Kwon~\etal \cite{kwon2016b}, better than 20\%. Good agreement with Kwon~\etal is also
seen in 29-like and 30-like, differing from our calculations by $\sim{20}$\% respectively. In contrast, consistently poor
agreement is observed between our data and Wu~\etal's \cite{wu2015a} for 28-like to 33-like and 35-like to 36-like, with differences
sometimes exceeding a factor $\sim{4}$. This includes ions for which we are in good agreement with Kwon~\etal.
It was found that agreement with Wu~\etal's 35-like results could be improved if core-rearrangement suppression was neglected 
from our inner-shell DR calculations.

We have assessed the effect of relativistic configuration mixing on partial DR rate coefficients in $n$ for 30-like
tungsten. This was done by comparing partial DR rate coefficients calculated in CA and IC. It was shown that the largest
difference between the two sets of DR rate coefficients at peak abundance temperature was a factor $\sim{2}$. This is a consequence
of the increasing complexity in atomic structure as we move along the isonuclear sequence. The increase in electrons available 
to promote increases the number of ``one-up, one-down" mixing configurations, hence increasing its importance. This result highlights
the importance of using IC DR results where ever possible.

For the purpose of calculating ionization fractions, we have also calculated the RR rate coefficients alongside the DR
rate coefficients in IC and CA. In IC, we included multipolar radiation contributions up to E40 and M39, whereas for CA
we only include multipolar contributions up to E40. For 19-like, the RR rate coefficients were found to contribute 14\%
to the recombination rate total at peak abundance temperature. This contribution decreases steadily with decreasing
residual charge to 4\% by 36-like.

Comparison of our recombination rate coefficients with the scaled data of P\"{u}tterich~\etal yielded relatively good
agreement for ions 19- to 26-like. Beyond 26-like we saw this agreement deteriorate rapidly as we moved along the
charge states. We also saw that agreement beyond 26-like was better when comparing our recombination rate coefficients
with P\"{u}tterich~\etal's unscaled data. However, the difference between our and P\"{u}tterich~\etal's data still exceeded
40\%. The large disagreement is to be expected due to the complexity of the ions being considered.

We have presented our DR rate coefficient calculations for ions W$^{55+}$ to W$^{38+}$. bringing us halfway into the ions
required to calculate the isonuclear sequence of tungsten. Unlike W$^{55+}$ to W$^{38+}$ considered in Preval~\etal \cite{preval2016a}, the
ions considered in this paper will be easily formed in ITER plasma conditions. Our next installment in {\it The Tungsten Project} will cover
the $4d^q$ ($q=1-10$) ions. We split the project in this way in anticipation of covering the $4f$-shell ions following
our $4d^q$ work.

\ack
SPP, NRB, and MGOM acknowledge the support of EPSRC grant EP/1021803
to the University of Strathclyde. All data calculated as part of this work are 
publicly available on the OPEN-ADAS website https://open.adas.ac.uk.

\section*{References}
\bibliography{simonpreval}
\newpage

\clearpage

\begin{table*}
\caption{Core-excitations $n-n'$ included in the DR rate coefficient calculations for W$^{55+}$ to W$^{38+}$. 
Core-excitations marked with an * were calculated only in CA, whilst the others were calculated in IC.}
\begin{tabular}{@{}llllll}
\hline
\hline
Ion-like & Symbol & Core-excitations & Ion & Symbol & Core-excitations \\
\hline
19-like   & W$^{55+}$  &  2--3$^{*}$, 3--3, 3--4, 3--5$^{*}$ & 28-like  & W$^{46+}$  &  3--4, 3--5$^{*}$           \\  
20-like   & W$^{54+}$  &  2--3$^{*}$, 3--3, 3--4, 3--5$^{*}$ & 29-like  & W$^{45+}$  &  3--4, 4--4, 4--5, 4--6$^{*}$ \\
21-like   & W$^{53+}$  &  2--3$^{*}$, 3--3, 3--4, 3--5$^{*}$ & 30-like  & W$^{44+}$  &  3--4, 4--4, 4--5, 4--6$^{*}$ \\
22-like   & W$^{52+}$  &  2--3$^{*}$, 3--3, 3--4, 3--5$^{*}$ & 31-like  & W$^{43+}$  &  3--4, 4--4, 4--5, 4--6$^{*}$ \\
23-like   & W$^{51+}$  &  2--3$^{*}$, 3--3, 3--4, 3--5$^{*}$ & 32-like  & W$^{42+}$  &  3--4, 4--4, 4--5, 4--6$^{*}$ \\
24-like   & W$^{50+}$  &  2--3$^{*}$, 3--3, 3--4, 3--5$^{*}$ & 33-like  & W$^{41+}$  &  3--4, 4--4, 4--5, 4--6$^{*}$ \\
25-like   & W$^{49+}$  &  2--3$^{*}$, 3--3, 3--4, 3--5$^{*}$ & 34-like  & W$^{40+}$  &  3--4, 4--4, 4--5, 4--6$^{*}$ \\
26-like   & W$^{48+}$  &  2--3$^{*}$, 3--3, 3--4, 3--5$^{*}$ & 35-like  & W$^{39+}$  &  3--4, 4--4, 4--5, 4--6$^{*}$ \\
27-like   & W$^{47+}$  &  2--3$^{*}$, 3--3, 3--4, 3--5$^{*}$ & 36-like  & W$^{38+}$  &  3--4, 4--4, 4--5, 4--6$^{*}$ \\
\hline
\hline
\label{table:drcorex}
\end{tabular}
\end{table*}

\begin{table}
\caption{Example of configurations included for DR rate coefficient calculation for the 30-like 4--5 core-excitation. 
Configurations marked with an * are included for mixing purposes by way of the one-up, one-down rule.}
\begin{tabular}{@{}lll}
\hline
\hline
$N$-electron & $(N+1)$-electron & \\
\hline
$4s^{2}$  &   $4s^{2} 5s$     &   $*4p 4d 5s$  \\
$4s 4p$   &   $4s^{2} 5p$     &   $*4p 4d 5p$  \\
$4s 4d$   &   $4s^{2} 5d$     &   $*4p 4d 5d$  \\
$4s 4f$   &   $4s^{2} 5f$     &   $*4p 4d 5f$  \\
$4s 5s$   &   $4s^{2} 5g$     &   $*4p 4d 5g$  \\
$4s 5p$   &   $4s 4p 5s$      &   $5s^{2}$  \\
$4s 5d$   &   $4s 4p 5p$      &   $5s 5p$  \\
$4s 5f$   &   $4s 4p 5d$      &   $5s 5d$  \\
$4s 5g$   &   $4s 4p 5f$      &   $5s 5f$  \\
$*4p^{2}$  &   $4s 4p 5g$      &   $5s 5g$  \\
$*4p 4d$   &   $4s 4d 5s$      &   $5p^{2}$  \\
          &   $4s 4d 5p$      &   $5p 5d$  \\
          &   $4s 4d 5d$      &   $5p 5f$  \\
          &   $4s 4d 5f$      &   $5p 5g$  \\
          &   $4s 4d 5g$      &   $5d^{2}$  \\
          &   $4s 4f 5s$      &   $5d 5f$  \\
          &   $4s 4f 5p$      &   $5d 5g$  \\
          &   $4s 4f 5d$      &   $5f^{2}$  \\
          &   $4s 4f 5f$      &   $5f 5g$  \\
          &   $4s 4f 5g$      &   $5g^{2}$  \\
          &   $*4p^{2} 5s$     &   \\
          &   $*4p^{2} 5p$     &   \\
          &   $*4p^{2} 5d$     &   \\
          &   $*4p^{2} 5f$     &   \\
          &   $*4p^{2} 5g$     &   \\
\hline
\hline
\label{table:drexam1}
\end{tabular}
\end{table}

\begin{table}
\caption{Example of configurations included for DR rate coefficient calculation for the 30-like 3--4
core-excitation. Configurations marked with an * are included for mixing purposes by way of the one-up, one-down
rule. Configurations with ``$n\ell$" are the core-rearrangement configurations. For example, the first core-rearrangement
configuration is formed by the reaction $3p^{5} 3d^{10} 4s^{2} 4p nl \rightarrow 3p^{6} 3d^{10} 4s n\ell$ + $e^{-}$.}
\begin{tabular}{@{}lll}
\hline
\hline
$N$-electron & $(N+1)$-electron & \\
\hline
$3p^{6} 3d^{10} 4s^{2}$      &  $3p^{6} 3d^{10} 4s^{2} 4p$      &  $3p^{5} 3d^{10} 4s^{2} 4p^{2}$  \\
$3p^{6} 3d^{10} 4s 4p $      &  $3p^{6} 3d^{10} 4s^{2} 4d$      &  $3p^{5} 3d^{10} 4s^{2} 4p 4d $  \\
$3p^{6} 3d^{10} 4s 4d $      &  $3p^{6} 3d^{10} 4s^{2} 4f$      &  $3p^{5} 3d^{10} 4s^{2} 4p 4f $  \\
$3p^{6} 3d^{10} 4s 4f $      &  $3p^{6} 3d^{10} 4s 4p^{2}$      &  $3p^{5} 3d^{10} 4s^{2} 4d^{2}$  \\
$*3p^{6} 3d^{10} 4p^{2}$     &  $3p^{6} 3d^{10} 4s 4p 4d $      &  $3p^{5} 3d^{10} 4s^{2} 4d 4f $  \\
$*3p^{6} 3d^{10} 4p 4d $     &  $*3p^{6} 3d^{10} 4p^{3}   $      &  $3p^{5} 3d^{10} 4s^{2} 4f^{2}$  \\
$3p^{6} 3d^{9}  4s^{2} 4p$   &  $*3p^{6} 3d^{10} 4p^{2} 4d$      &  $*3p^{5} 3d^{10} 4s 4p^{3}    $  \\
$3p^{6} 3d^{9}  4s^{2} 4d$   &  $*3p^{6} 3d^{10} 4p^{2} 4f$      &  $*3p^{5} 3d^{10} 4s 4p^{2} 4d $  \\
$3p^{6} 3d^{9}  4s^{2} 4f$   &  $*3p^{6} 3d^{10} 4p 4d^{2}$      &  $*3p^{5} 3d^{10} 4s 4p^{2} 4f $  \\
$*3p^{6} 3d^{9}  4s 4p^{2}$  &  $*3p^{6} 3d^{10} 4p 4d 4f $      &  $*3p^{5} 3d^{10} 4s 4p 4d^{2} $  \\
$*3p^{6} 3d^{9}  4s 4p 4d $  &  $3p^{6} 3d^{10} 4s 4p 4f $      &  $*3p^{5} 3d^{10} 4s 4p 4d 4f  $  \\
$3p^{5} 3d^{10} 4s^{2} 4p$   &  $3p^{6} 3d^{10} 4s 4d^{2}$      &  \\
$3p^{5} 3d^{10} 4s^{2} 4d$   &  $3p^{6} 3d^{10} 4s 4d 4f $      &  \\
$3p^{5} 3d^{10} 4s^{2} 4f$   &  $3p^{6} 3d^{10} 4s 4f^{2}$      &  \\
$*3p^{5} 3d^{10} 4s 4p^{2}$  &  $3p^{6} 3d^{9} 4s^{2} 4p^{2}$   &  \\
$*3p^{5} 3d^{10} 4s 4p 4d $  &  $3p^{6} 3d^{9} 4s^{2} 4p 4d $   &  \\
$3p^{6} 3d^{10} 4p 4f $      &  $3p^{6} 3d^{9} 4s^{2} 4p 4f $   &  \\
$3p^{6} 3d^{10} 4d^{2}$      &  $3p^{6} 3d^{9} 4s^{2} 4d^{2}$   &  \\
$3p^{6} 3d^{10} 4d 4f $      &  $3p^{6} 3d^{9} 4s^{2} 4d 4f $   &  \\
$3p^{6} 3d^{10} 4f^{2}$      &  $3p^{6} 3d^{9} 4s^{2} 4f^{2}$   &  \\
$3p^{6} 3d^{10} 4s n\ell$    &  $*3p^{6} 3d^{9} 4s 4p^{3}    $   &  \\
$3p^{6} 3d^{10} 4p n\ell$    &  $*3p^{6} 3d^{9} 4s 4p^{2} 4d $   &  \\
$3p^{6} 3d^{10} 4d n\ell$    &  $*3p^{6} 3d^{9} 4s 4p^{2} 4f $   &  \\
$3p^{6} 3d^{10} 4f n\ell$    &  $*3p^{6} 3d^{9} 4s 4p 4d^{2} $   &  \\
                             &  $*3p^{6} 3d^{9} 4s 4p 4d 4f  $   &  \\
\hline
\hline
\label{table:drexam2}
\end{tabular}
\end{table}

\begin{table}
\caption{Example of configurations included for RR rate coefficient calculation for 30-like. Configurations marked with an *
are included for mixing purpose by way of the one-up, one-down rule.}
\begin{tabular}{@{}ll}
\hline
\hline
$N$-electron & $(N+1)$-electron \\
\hline
$4s^{2}$  & $4s^{2} 4p$ \\
$4s 4p$   & $4s^{2} 4d$ \\
$4s 4d$   & $4s^{2} 4f$ \\
$4s 4f$   & $4s 4p^{2}$ \\
$*4p 4d$  & $4s 4p 4d$ \\
$*4p^{2}$ & $4s 4p 4f$ \\
          & $4s 4d^{2}$ \\
          & $4s 4d 4f$ \\
          & $4s 4f^{2}$ \\
          & $*4p^{2} 4d$ \\
          & $*4p 4d^{2}$ \\
          & $*4p 4d 4f$ \\
          & $*4p^{3}$ \\
          & $*4p^{2} 4f$ \\
\hline
\hline
\label{table:rrexam}
\end{tabular}
\end{table}

\begin{table}
\caption{Comparison of 27-like total DR rate coefficients between the current work, and Meng~\etal \cite{meng2009a}. Note $[x]=10^{x}$.}
\begin{tabular}{@{}llll}
\hline
\hline
Temp (K) & This work & Meng & \% Diff \\
\hline
$4.70[+5]$ & $9.47[-10]$ & $1.04[-09]$ & $9.63$ \\
$6.37[+5]$ & $9.05[-10]$ & $9.70[-10]$ & $7.14$ \\
$8.64[+5]$ & $8.48[-10]$ & $8.89[-10]$ & $4.87$ \\
$1.17[+6]$ & $7.70[-10]$ & $7.99[-10]$ & $3.74$ \\
$1.59[+6]$ & $6.75[-10]$ & $6.99[-10]$ & $3.50$ \\
$2.16[+6]$ & $5.75[-10]$ & $5.99[-10]$ & $4.09$ \\
$2.92[+6]$ & $4.81[-10]$ & $5.12[-10]$ & $6.52$ \\
$3.97[+6]$ & $4.00[-10]$ & $4.43[-10]$ & $10.8$ \\
$5.38[+6]$ & $3.36[-10]$ & $3.89[-10]$ & $15.8$ \\
$7.30[+6]$ & $2.85[-10]$ & $3.44[-10]$ & $20.7$ \\
$9.90[+6]$ & $2.39[-10]$ & $2.98[-10]$ & $24.7$ \\
$1.34[+7]$ & $1.95[-10]$ & $2.49[-10]$ & $27.6$ \\
$1.82[+7]$ & $1.53[-10]$ & $1.98[-10]$ & $29.2$ \\
$2.47[+7]$ & $1.16[-10]$ & $1.51[-10]$ & $29.9$ \\
$3.35[+7]$ & $8.44[-11]$ & $1.10[-10]$ & $30.4$ \\
$4.54[+7]$ & $5.94[-11]$ & $7.77[-11]$ & $30.7$ \\
$6.16[+7]$ & $4.08[-11]$ & $5.34[-11]$ & $31.1$ \\
$8.35[+7]$ & $2.74[-11]$ & $3.60[-11]$ & $31.3$ \\
$1.13[+8]$ & $1.82[-11]$ & $2.39[-11]$ & $31.2$ \\
$1.54[+8]$ & $1.19[-11]$ & $1.56[-11]$ & $31.0$ \\
$2.08[+8]$ & $7.78[-12]$ & $1.02[-11]$ & $30.6$ \\
$2.83[+8]$ & $5.03[-12]$ & $6.55[-12]$ & $30.3$ \\
$3.83[+8]$ & $3.23[-12]$ & $4.21[-12]$ & $30.2$ \\
$5.20[+8]$ & $2.07[-12]$ & $2.69[-12]$ & $30.4$ \\
$7.05[+8]$ & $1.31[-12]$ & $1.72[-12]$ & $30.9$ \\
\hline
\hline
\label{table:mengrates}
\end{tabular}
\end{table}

\begin{table*}
\centering
\caption{Table of differences between the total DR rate coefficients as calculated in this work, and by Wu~\etal \cite{wu2015a}. The \% difference between the two DR rate coefficients is given relative to the current work. Note
$[x]=10^{x}$.}
\begin{tabular}{@{}cccccc}
\hline
\hline
Temp (K) & \% Diff 31 & \% Diff 32 & \% Diff 33 & \% Diff 35 & \% Diff 36 \\ 
\hline
$5.80[+4]$ & $ 13.2$ & $-38.3$ & $-51.1$ & $-63.2$ & $-64.4$ \\
$8.71[+4]$ & $ 16.2$ & $-36.1$ & $-50.1$ & $-61.4$ & $-56.0$ \\
$1.31[+5]$ & $ 7.70$ & $-36.7$ & $-49.1$ & $-60.5$ & $-48.2$ \\
$1.96[+5]$ & $ 0.23$ & $-36.5$ & $-47.5$ & $-58.3$ & $-42.1$ \\
$2.95[+5]$ & $-3.11$ & $-35.9$ & $-45.9$ & $-53.5$ & $-37.5$ \\
$4.43[+5]$ & $-4.88$ & $-35.4$ & $-45.0$ & $-46.0$ & $-32.3$ \\
$6.65[+5]$ & $-5.90$ & $-35.9$ & $-45.5$ & $-36.9$ & $-25.9$ \\
$9.98[+5]$ & $-6.23$ & $-37.2$ & $-46.5$ & $-28.5$ & $-18.8$ \\
$1.50[+6]$ & $-5.95$ & $-37.4$ & $-45.8$ & $-21.5$ & $-11.1$ \\
$2.25[+6]$ & $-2.66$ & $-35.2$ & $-41.9$ & $-13.4$ & $-0.75$ \\
$3.38[+6]$ & $ 4.54$ & $-29.4$ & $-33.8$ & $-2.48$ & $ 13.4$ \\
$5.07[+6]$ & $ 15.2$ & $-18.6$ & $-21.2$ & $ 12.8$ & $ 32.2$ \\
$7.62[+6]$ & $ 30.6$ & $-2.43$ & $-5.09$ & $ 33.1$ & $ 56.1$ \\
$1.14[+7]$ & $ 49.1$ & $ 17.3$ & $ 12.3$ & $ 57.1$ & $ 83.6$ \\
$1.72[+7]$ & $ 66.5$ & $ 37.0$ & $ 28.5$ & $ 81.6$ & $ 110 $ \\
$2.58[+7]$ & $ 80.4$ & $ 53.9$ & $ 42.5$ & $ 104 $ & $ 133 $ \\
$3.87[+7]$ & $ 90.6$ & $ 67.1$ & $ 53.5$ & $ 121 $ & $ 152 $ \\
$5.81[+7]$ & $ 98.5$ & $ 76.9$ & $ 61.7$ & $ 135 $ & $ 166 $ \\
$8.73[+7]$ & $ 104 $ & $ 84.2$ & $ 67.4$ & $ 145 $ & $ 178 $ \\
$1.31[+8]$ & $ 107 $ & $ 89.8$ & $ 71.2$ & $ 152 $ & $ 186 $ \\
$1.97[+8]$ & $ 109 $ & $ 93.7$ & $ 73.7$ & $ 157 $ & $ 190 $ \\
$2.95[+8]$ & $ 111 $ & $ 95.7$ & $ 75.3$ & $ 160 $ & $ 192 $ \\
$4.44[+8]$ & $ 112 $ & $ 96.9$ & $ 76.6$ & $ 162 $ & $ 195 $ \\
$6.66[+8]$ & $ 113 $ & $ 97.8$ & $ 77.7$ & $ 164 $ & $ 197 $ \\
$1.00[+9]$ & $ 113 $ & $ 98.7$ & $ 78.2$ & $ 165 $ & $ 199 $ \\
\hline
\hline
\label{table:wutable}
\end{tabular}
\end{table*}

\begin{table*}
\caption{Comparison of peak abundance temperatures and fractions as calculated using P\"{u}tterich~\etal's data \cite{putterich2008a}, and
P\"{u}tterich~\etal's data with 01- to 36-like replaced with our data. The ionization rate coefficients originate from Loch~\etal \cite{loch2005a}. Note $[x]=10^{x}$.}
\begin{tabular}{@{}lccccccc}
\hline
\hline
Ion-like & Charge & Putt $T_{\mathrm{peak}}$ & Putt $f_{\mathrm{peak}}$ & This work $T_{\mathrm{peak}}$ & This work $f_{\mathrm{peak}}$ & $\Delta{T}$\% & $\Delta{f}$\% \\
\hline
01-like & W$^{73+}$ & $3.88[+9]$ & $0.440$ & $4.06[+9]$ & $0.426$ & $-4.50$ & $ 3.14$ \\
02-like & W$^{72+}$ & $2.29[+9]$ & $0.442$ & $2.46[+9]$ & $0.405$ & $-7.11$ & $ 9.16$ \\
03-like & W$^{71+}$ & $1.48[+9]$ & $0.360$ & $1.64[+9]$ & $0.363$ & $-10.1$ & $-0.93$ \\
04-like & W$^{70+}$ & $9.66[+8]$ & $0.294$ & $1.06[+9]$ & $0.304$ & $-8.84$ & $-3.30$ \\
05-like & W$^{69+}$ & $7.17[+8]$ & $0.247$ & $7.67[+8]$ & $0.255$ & $-6.52$ & $-3.19$ \\
06-like & W$^{68+}$ & $5.72[+8]$ & $0.218$ & $5.97[+8]$ & $0.228$ & $-4.19$ & $-4.17$ \\
07-like & W$^{67+}$ & $4.76[+8]$ & $0.199$ & $4.84[+8]$ & $0.213$ & $-1.61$ & $-6.28$ \\
08-like & W$^{66+}$ & $4.03[+8]$ & $0.189$ & $3.98[+8]$ & $0.209$ & $ 1.36$ & $-9.33$ \\
09-like & W$^{65+}$ & $3.46[+8]$ & $0.195$ & $3.32[+8]$ & $0.217$ & $ 3.94$ & $-9.99$ \\
10-like & W$^{64+}$ & $2.99[+8]$ & $0.214$ & $2.82[+8]$ & $0.229$ & $ 6.25$ & $-6.57$ \\
11-like & W$^{63+}$ & $2.60[+8]$ & $0.202$ & $2.43[+8]$ & $0.198$ & $ 7.04$ & $ 1.78$ \\
12-like & W$^{62+}$ & $2.25[+8]$ & $0.188$ & $2.11[+8]$ & $0.179$ & $ 6.48$ & $ 5.25$ \\
13-like & W$^{61+}$ & $2.01[+8]$ & $0.182$ & $1.89[+8]$ & $0.174$ & $ 6.19$ & $ 4.53$ \\
14-like & W$^{60+}$ & $1.80[+8]$ & $0.172$ & $1.70[+8]$ & $0.171$ & $ 6.25$ & $ 1.01$ \\
15-like & W$^{59+}$ & $1.62[+8]$ & $0.163$ & $1.52[+8]$ & $0.166$ & $ 6.50$ & $-1.35$ \\
16-like & W$^{58+}$ & $1.47[+8]$ & $0.157$ & $1.38[+8]$ & $0.167$ & $ 6.74$ & $-6.02$ \\
17-like & W$^{57+}$ & $1.36[+8]$ & $0.139$ & $1.27[+8]$ & $0.154$ & $ 7.02$ & $-9.48$ \\
18-like & W$^{56+}$ & $1.26[+8]$ & $0.140$ & $1.16[+8]$ & $0.157$ & $ 7.99$ & $-10.8$ \\
19-like & W$^{55+}$ & $1.17[+8]$ & $0.147$ & $1.07[+8]$ & $0.150$ & $ 8.96$ & $-1.73$ \\
20-like & W$^{54+}$ & $1.08[+8]$ & $0.150$ & $9.83[+7]$ & $0.160$ & $ 10.2$ & $-6.24$ \\
21-like & W$^{53+}$ & $1.01[+8]$ & $0.148$ & $9.01[+7]$ & $0.162$ & $ 11.6$ & $-9.00$ \\
22-like & W$^{52+}$ & $9.29[+7]$ & $0.144$ & $8.33[+7]$ & $0.167$ & $ 11.6$ & $-13.6$ \\
23-like & W$^{51+}$ & $8.60[+7]$ & $0.143$ & $7.73[+7]$ & $0.153$ & $ 11.3$ & $-6.36$ \\
24-like & W$^{50+}$ & $8.01[+7]$ & $0.146$ & $7.14[+7]$ & $0.156$ & $ 12.2$ & $-5.99$ \\
25-like & W$^{49+}$ & $7.41[+7]$ & $0.159$ & $6.57[+7]$ & $0.164$ & $ 12.7$ & $-2.96$ \\
26-like & W$^{48+}$ & $6.74[+7]$ & $0.181$ & $6.02[+7]$ & $0.174$ & $ 12.1$ & $ 4.26$ \\
27-like & W$^{47+}$ & $5.97[+7]$ & $0.136$ & $5.47[+7]$ & $0.182$ & $ 9.14$ & $-25.2$ \\
28-like & W$^{46+}$ & $5.10[+7]$ & $0.353$ & $5.03[+7]$ & $0.188$ & $ 1.45$ & $ 88.0$ \\
29-like & W$^{45+}$ & $4.32[+7]$ & $0.227$ & $4.62[+7]$ & $0.157$ & $-6.68$ & $ 43.9$ \\
30-like & W$^{44+}$ & $3.74[+7]$ & $0.196$ & $4.21[+7]$ & $0.150$ & $-11.2$ & $ 30.8$ \\
31-like & W$^{43+}$ & $3.40[+7]$ & $0.194$ & $3.80[+7]$ & $0.177$ & $-10.4$ & $ 10.1$ \\
32-like & W$^{42+}$ & $3.22[+7]$ & $0.108$ & $3.34[+7]$ & $0.244$ & $-3.56$ & $-55.8$ \\
33-like & W$^{41+}$ & $3.11[+7]$ & $0.044$ & $3.06[+7]$ & $0.210$ & $ 1.64$ & $-79.1$ \\
34-like & W$^{40+}$ & $3.01[+7]$ & $0.052$ & $2.84[+7]$ & $0.149$ & $ 5.80$ & $-65.2$ \\
35-like & W$^{39+}$ & $2.91[+7]$ & $0.049$ & $2.67[+7]$ & $0.108$ & $ 9.18$ & $-54.6$ \\
36-like & W$^{38+}$ & $2.81[+7]$ & $0.093$ & $2.52[+7]$ & $0.104$ & $ 11.4$ & $-10.3$ \\
\hline
\hline
\label{table:peaktemp}
\end{tabular}
\end{table*}

\clearpage

\end{document}